\documentclass[11pt,a4paper]{article}
\usepackage[T1]{fontenc}
\usepackage[utf8]{inputenc}
\usepackage{lmodern}
\usepackage[margin=25mm]{geometry}
\usepackage{microtype}
\usepackage{amsmath,amssymb,amsthm,mathtools}
\usepackage{stmaryrd}
\usepackage{enumitem,xcolor}
\usepackage[colorlinks=true,linkcolor=blue!45!black,citecolor=blue!45!black,urlcolor=blue!45!black]{hyperref}
\usepackage[nameinlink,noabbrev]{cleveref}
\usepackage{booktabs,array,needspace}

\setlist{itemsep=3pt,topsep=5pt}
\newtheorem{theorem}{Theorem}[section]
\newtheorem{lemma}[theorem]{Lemma}
\newtheorem{proposition}[theorem]{Proposition}
\newtheorem{corollary}[theorem]{Corollary}
\theoremstyle{definition}
\newtheorem{definition}[theorem]{Definition}
\newtheorem{example}[theorem]{Example}
\theoremstyle{remark}
\newtheorem{remark}[theorem]{Remark}
\newcommand{\LFD}{\mathsf{LFD}}
\newcommand{\LI}{\mathsf{LFD}+\mathsf I}
\newcommand{\LC}{\mathsf{LFD}+\mathsf{Con}}
\newcommand{\LE}{\mathsf{LFD}^{=}}
\newcommand{\FO}{\mathsf{FO}}
\newcommand{\Dmod}{\mathbb D}
\newcommand{\Emod}{\mathsf E}
\newcommand{\All}{\mathsf A}
\newcommand{\ST}{\operatorname{ST}}
\newcommand{\Copy}{\operatorname{Copy}}
\newcommand{\Eq}{\operatorname{Eq}}
\newcommand{\sat}{\models}
\newcommand{\N}{\mathbb{N}}
\newcommand{\M}{\mathfrak M}
\newcommand{\B}{\mathfrak B}
\newcommand{\one}{\operatorname{one}}
\newcommand{\Dia}[1]{\Emod_{\{#1\}}}
\newcommand{\Boxx}[1]{\Dmod_{\{#1\}}}
\DeclareMathOperator{\SAT}{SAT}

\newcommand{\Lists}{\operatorname{List}}
\newcommand{\Hd}{\operatorname{head}}
\newcommand{\Tl}{\operatorname{tail}}
\newcommand{\Eval}{\operatorname{Eval}}
\newcommand{\Save}{\mathsf{Save}}
\newcommand{\Write}{\mathsf{Write}}
\newcommand{\Read}{\mathsf{Read}}
\newcommand{\Seed}{\mathsf{Seed}}
\newcommand{\Load}{\mathsf{Load}}
\newcommand{\Lookup}{\mathsf{Lookup}}
\newcommand{\PushPtr}{\mathsf{PushPtr}}
\newcommand{\PopPtr}{\mathsf{PopPtr}}

\title{\textbf{Connecting Extended Modal Dependence Logics}}

\author{Johan van Benthem, Chenwei Shi and Wenfei Ouyang}
\begin{document}
\maketitle
\begin{abstract} We consider three natural extensions of the basic decidable dependence logic $\mathsf{LFD}$, adding, respectively, conditional dependence atoms, independence atoms, and identity atoms between variables. We show that these systems are all connected to first-order logic with identity by SAT reductions and in several cases also faithful translations. In particular, this implies that all these extended modal-style dependence logics are undecidable. This solves a number of problems left open in the original publication \cite{BV21} introducing the $\mathsf{LFD}$ framework. The definitions and proofs presented here were found by $\mathsf{ChatGPT}$ and checked and streamlined as needed by the authors. At the end, we provide some discussion of what new things we  learnt about dependence logics and about the methodology of using AI in logical research.

\end{abstract}

\section{Languages, dependence semantics, and the results}

We work over nonempty object domains, finite variable sets, and finite
relational vocabularies. First-order logic has equality but no function
or constant symbols. Vocabularies and variable sets may grow effectively
with input formulas in the reductions to follow. Nullary predicates, if present, are carried along
as truth values; the nontrivial predicate constructions in what follows below concern
positive arities.

\smallskip

{\bf The basic system.} A dependence model is a tuple of the form $\mathcal M=(\mathfrak M,A)$, where $\mathfrak M$
is a relational structure with domain $U\ne\varnothing$, and
$\varnothing\ne A\subseteq U^V$ is a team of admissible assignments.
For $s,t\in A$ and $X\subseteq V$, write
\[
 s=_Xt \quad\Longleftrightarrow\quad
 s(x)=t(x)\text{ for every }x\in X.
\]
The language of the basic dependence logic $\mathsf{LFD}$ of \cite{BV21} has standard relational atoms and Boolean connectives which receive their ordinary 
semantics at assignments in dependence models. The  essential semantic clauses for $\mathsf{LFD}$ concern its  dependence modality and dependence atoms:
\begin{align}
 \mathcal M,s\sat\Dmod_X\psi
 &\iff \forall t\in A\,(s=_Xt\Rightarrow\mathcal M,t\sat\psi),
 \label{eq:box}\\
 \mathcal M,s\sat D_Xy
 &\iff \forall t\in A\,(s=_Xt\Rightarrow s(y)=t(y)).
 \label{eq:dep}
\end{align}
Define $\Emod_X\psi:=\neg\Dmod_X\neg\psi$ and
$\All\psi:=\Dmod_\varnothing\psi$. Thus $\Emod_X$ existentially varies
an assignment while fixing $X$, and $\All$ quantifies over all variables in the whole language considered, which has the effect of a global universal modality over all assignments present  in the model.

\smallskip

{\bf Three extensions.} There are several natural extensions for the above basic language.  

The language $\LI$ adds the basic local independence atoms of
\cite[Section~6.3]{BV21}:
\begin{equation}
 \mathcal M,s\sat I_Xy
 \iff
 \forall t\in A\;\exists u\in A\,
       (u=_Xs\land u(y)=t(y)).
 \label{eq:ind}
\end{equation}
%Only a single dependent-side variable $y$ is needed in our construction.
The language $\LC$ adds source-relative conditional dependence:
\begin{equation}
 \mathcal M,s\sat D_X^\alpha y
 \iff
 \forall t\in A\,
  (s=_Xt\land\mathcal M,t\sat\alpha\Rightarrow s(y)=t(y)).
 \label{eq:cond}
\end{equation}
The current assignment $s$ need not satisfy $\alpha$. Conditions may be
arbitrary formulas, including nested conditional dependences, and are
evaluated in the original model. This is the conditional operator in
\cite[Section~6.4]{BV21}. 

\begin{remark} There are also variant definitions of conditional dependence atoms, for instance, requiring the vantage point $s$ itself to satisfy the condition $\alpha$. We ignore these here.  \end{remark}

Finally, the system $\LE$ adds actual equality atoms:
$\mathcal M,s\sat a=b$ iff $s(a)=s(b)$.

\smallskip

We now state the central notions of system connection involved in the main results to follow.

\smallskip

{\bf SAT reductions.} Our first notion establishes an effective connection between the satisfiability problems for two logics $L_1, L_2$, operating at the level of their computational complexity.

\begin{definition}
A \emph{satisfiability reduction} from $L_1$ to $L_2$ is a
computable map $f$ on formulas satisfying the following equivalence:

\vspace{-4ex}

\begin{align*}
 \varphi\in\SAT(L_1)&\iff f(\varphi)\in\SAT(L_2).
% \varphi\in\FinSAT(L_1)&\iff f(\varphi)\in\FinSAT(L_2).
\end{align*}
For dependence logics, satisfiability means truth at some $s\in A$.
%Finite satisfiability requires a finite object domain. Since the variable set is finite, its team is then finite as well. 
For first-order logic we use satisfiability of formulas under some assignment of objects to their free variables.
\end{definition}

\vspace{-1ex}

Next we consider a stronger connection between arbitrary logical systems $L_1$ and $L_2$.

\smallskip

{\bf Translations.}  A translation from $L_1$ into $L_2$ comes in two parts, a syntactic map ${\bf t}$ taking formulas of $L_1$ to formulas of $L_2$, and a semantic transformation ${\bf T}$ taking models for $L_2$ to models for $L_1$. These are required to satisfy the following basic contravariant equivalence for all models $\mathfrak{M}$ for $L_2$ and formulas $\varphi$ of $L_1$:\footnote{We  suppress assignments to free variables for now: our later results will make it clear how these are handled.}

\vspace{-1ex}

\begin{quote}

 {\bf T}($\mathfrak{M})\models \varphi$ \quad iff \quad $\mathfrak{M}\models$ {\bf t}($\varphi$). 

\end{quote}

\vspace{-1ex}

If the transformation {\bf T} is surjective, this will immediately yield the usual equivalence of the two assertions $\varphi_1, \dots, \varphi_k \models \psi$ and  {\bf t}$(\varphi_1), \dots, $ {\bf t}$(\varphi_k)  \models$ {\bf t}$( \psi)$. However, the transformation {\bf T} may also have a more restricted domain, as will be the case in many results to follow,  defined by a condition $\Gamma$ in the language of $L_2$. In that case, the reduction equivalence becomes the format found in several later theorems:

\vspace{-1ex}

\begin{quote}

$\varphi_1, \dots, \varphi_k \models \psi$ \quad iff \quad  $\Gamma$, {\bf t}$(\varphi_1), \dots, $ {\bf t}$(\varphi_k)  \models$ {\bf t}$( \psi)$\footnote{There are sometimes ways of `hiding' various types of condition $\Gamma$ inside the syntactic language translation {\bf t}, but we will not go into these here.}

\end{quote}

\vspace{-1ex}

We will call translations satisfying this reduction of consequence {\it faithful}. 

\begin{remark}[A weaker notion] In Section 3, we will also encounter a weaker notion of translation which only satisfies part of the above contra-variant equivalence. Also, the term `translation' sometimes only refers to a syntactic mapping, a narrower usage that is found occasionally below. \end{remark}

\begin{remark}[The fine-print: fragments] The preceding presentation does not fit the proofs of later theorems exactly. Usually, these proofs analyze only some fragment of the relevant logical languages with a restricted number of variables and atomic predicates, and the above connection then only applies for models of the right similarity type for these fragments. The relevant conditions $\Gamma$ will also depend on these finite vocabularies. Concrete examples can be found in the various sections of this Note. The results for the complete languages then follows by `patching' these fragment-based results in a way that we will not spell out in this note.\end{remark}   

In particular, if the syntactic mapping {\bf t}  in a translation like this is effective, we get a conservative  SAT reduction from the logic $L_1$ into the logic $L_2$ (provided the condition $\Gamma$ is finite), which helps us determine decidability or undecidability of logics. But translations in our sense can also transfer further properties of logical systems, cf. \cite{JB26}, \cite{BCK22}.

\vspace{1ex}

{\bf Main results.} We have found a number of new, and sometimes quite surprising, connections between the three extended modal dependence logics considered here and first-order logic with identity $\mathsf{FO(=)}$, the central working system of classical logic. Before explaining these, we start with a number of standard observations explaining the centrality of the system $\mathsf{FO(=)}$.

\smallskip

{\bf Standard translations into $\mathsf{FO(=)}$.} For a start, we state a result which immediately provides one direction for all reductions considered in this Note. It is well-known that all our logics are SAT-reducible to first-order logic with identity by providing `standard translations' which essentially reproduce the above semantics in first-order language.

Fix an enumeration $V=\{v_1,\ldots,v_m\}$ and introduce a fresh $m$-ary
relation symbol $T$ for the team. We may add an unused variable when
necessary to ensure $m\ge1$. Vectors $\bar s,\bar t,\bar u$ below are
vectors of first-order variables, not named assignments in the target
language. Put
\[
 \operatorname{agr}_X(\bar s,\bar t)
 :=\bigwedge_{v_i\in X}s_i=t_i.
\]
An empty conjunction is $\top$. If $y=v_j$, write $s_y:=s_j$.

Define a standard translation at a current vector $\bar s$ by
\begin{align}
 \ST_{\bar s}(P(a_1,\ldots,a_r))
 &:=P(s_{a_1},\ldots,s_{a_r}),\notag\\
 \ST_{\bar s}(a=b)&:=s_a=s_b,\notag\\
 \ST_{\bar s}(\Dmod_X\psi)
 &:=\forall\bar t\bigl(
   (T(\bar t)\land\operatorname{agr}_X(\bar s,\bar t))
       \to\ST_{\bar t}(\psi)\bigr),\notag\\
 \ST_{\bar s}(D_Xy)
 &:=\forall\bar t\bigl(
   (T(\bar t)\land\operatorname{agr}_X(\bar s,\bar t))
       \to s_y=t_y\bigr),\notag\\
 \ST_{\bar s}(I_Xy)
 &:=\forall\bar t\Bigl(T(\bar t)\to
     \exists\bar u\bigl(T(\bar u)\land
       \operatorname{agr}_X(\bar s,\bar u)\land u_y=t_y\bigr)\Bigr),
 \label{eq:st-ind}\\
 \ST_{\bar s}(D_X^\alpha y)
 &:=\forall\bar t\Bigl(
     \bigl(T(\bar t)\land\operatorname{agr}_X(\bar s,\bar t)
                   \land\ST_{\bar t}(\alpha)\bigr)
       \to s_y=t_y\Bigr).
 \label{eq:st-cond}
\end{align}
The equality clause is used only for $\LE$. The translation commutes
with Boolean connectives. Bound vectors are chosen fresh at every step,
so there is no variable capture. Define the sentence
\begin{equation}
 S_L(\varphi):=\exists\bar s\,
          (T(\bar s)\land\ST_{\bar s}(\varphi)),
 \qquad L\in\{\LI,\LC,\LE\}.
 \label{eq:st-sentence}
\end{equation}

\begin{proposition}
\label{prop:st}
Each $S_L$ is a faithful translation of $L$ into $\FO(=)$.
\end{proposition}

\begin{proof}
Given $\mathcal M=(\mathfrak M,A)$, expand $\mathfrak M$ by
\[
 T^{\mathfrak M_A}
 :=\{(s(v_1),\ldots,s(v_m)):s\in A\}.
\]
Structural induction gives, for every $s\in A$,
\[
 \mathcal M,s\sat\varphi
 \iff
 \mathfrak M_A\sat\ST_{\bar s}(\varphi)[\bar s\mapsto s(V)].
\]
The two new cases are exactly the quantified conditions in
\eqref{eq:ind} and \eqref{eq:cond}. In the conditional case the induction
hypothesis applies to $\alpha$ at each $t\in A$, so nesting causes no
additional difficulty.

Conversely, any model of \eqref{eq:st-sentence} determines the nonempty
team consisting of its $T$-tuples. The same induction gives a satisfying
point of the corresponding dependence model. Both constructions keep
the object domain unchanged and therefore preserve finiteness.
\end{proof}

\vspace{-1ex}

{\bf Less immediate system connections.} Next come the  new results of this Note. Section 2 presents a faithful translation from $\mathsf{FO(=)}$ into $\LI$, extending and correcting a result in \cite{BV21}. Section 3 presents a SAT reduction from $\mathsf{FO}$ into $\LE$ extending the known result of \cite{GP21} for the Kahr class, the $\forall \exists \forall$-type prenex fragment of first-order logic with an undecidable satisfiability problem. Next, Section 4 presents a faithful translation from $\LE$ into $\LC$ which stays close to the actual structure of the relevant logical formulas. Composed with the preceding result this gives us a SAT reduction from $\mathsf{FO(=)}$ into $\LC$. However, we have not been able to find a direct faithful translation from  $\mathsf{FO(=)}$ into $\LC$. Instead, Appendix A gives an indirect SAT reduction from $\mathsf{FO(=)}$ into $\LC$ by encoding a geometric Tiling Problem having the same  complexity as the satisfiability problem for  $\mathsf{FO(=)}$. 

Taken together and composing arrows, these results show us that all three extended dependence logics considered in this Note are mutually SAT reducible among each other and also with first-order logic. In particular then, all these systems are undecidable. What is left open by our results is whether each of these SAT reductions can be replaced by a faithful translation.\footnote{Even logics that are mutually faithfully translatable are not equivalent by definition. For counterexamples to direct definability for each pair of our extended dependence logics, see Appendix B.}

\smallskip

{\bf Consequences and further directions.}  The non-routine results in this Note all use new model constructions, and bring to light some surprising strengths of our modal dependence logics. These include the remarkable power of conditional functional dependence (originally taken to be a routine extension of the non-conditional case) and the surprising effects of adding identity which makes dependence models a powerful computational device. We will briefly list some of these conceptual insights in Section 5 below which reflects in various ways on the results presented in this Note. We also state our experiences with the new AI tools for research in logic, in a dynamic of asking questions
and getting to know our existing systems ever better.

\smallskip
{\bf Appendices.} As stated already, Appendix A shows how to effectively encode a geometric tiling benchmark task as a satisfiability problem in $\LC$. Appendix B shows how none of our three extended dependence logics are directly definable in others. And finally, Appendix C briefly summarizes our interactions with AI which led to the present Note.

\smallskip

%For a start, noting that conditional dependence atoms can also be sees as adding a new modality to $\mathsf{LFD}$, we conclude that the `Boolean Dependence Logic' of \textcolor{red}{{\bf CHENWEI ET AL.}} which extends $\mathsf{LFD}$ is undecidable.

%We also identify the main conceptual insights embodied in our translations, which include 

%We briefly explain our interactions with AI in Appendix C. Referring to that history, 
%\medskip
%
{\bf Final note.} This paper is not put on $\mathsf{arXiv}$ to claim priority or conform to other old publication habits. We see it rather as a small case study of what can be achieved when human researchers work with AI systems to expand their problem-solving power and to extend their scholarly grasp of the relevant literature from the vast corpus produced by the logical community. As for the mathematical proofs presented here, we have checked them and they make sense to us, but this posting now makes them available for scrutiny by a broader audience.

\section{Translating $\mathsf{FO(=)}$ into $\LI$ } 

In this section we provide a faithful translation from first-order logic with identity into the logic $\LI$ of dependence and independence. We will concentrate on arbitrary models, though, inspecting the text, our constructions and arguments also apply to correlating finite models.

\vspace{-1ex}

\subsection{The main result}

\begin{theorem}%[Main result]
\label{thm:main}
There is a faithful translation from  $\mathsf{FO(=)}$ into the dependence logic $\LI$.\footnote{We have changed the original result obtained by $\mathsf{ChatGPT}$, namely an effective SAT reduction, to a faithful translation, since this stronger result is in effect what the AI-generated proof established.} 
%More precisely,
%\begin{align*}
% \SAT(\LI)&\equiv_m\SAT(\FO(=)),\\
% \FinSAT(\LI)&\equiv_m\FinSAT(\FO(=)).
%\end{align*}
%The two equivalences are witnessed by the same formula translations.
\end{theorem}

%The proof uses the following routes, with every arrow denoting a
%conservative satisfiability reduction:
%\begin{align}
% \LI&\longrightarrow\FO\longrightarrow\LE\longrightarrow\LC,
% \label{eq:route-one}\\
% \LC&\longrightarrow\FO\longrightarrow\LI.
% \label{eq:route-two}
%\end{align}
%We establish all arrows explicitly except $\FO\longrightarrow\LE$,
%for which we state the published theorem precisely in
%\cref{sec:composition}. Mere undecidability of two logics would not
%justify either route.
%
%\section{Translations into first-order logic}
%\label{sec:st}

%\section{From first-order logic to local independence}
%\label{sec:fo-ind}

Let $\chi$ be a relational first-order sentence in first-order logic with identity. Let $V$ contain all
variable names occurring in $\chi$, adding an unused variable if needed.
Set
\[
 k:=\max\bigl(\{1\}\cup
       \{\operatorname{arity}(P):P\text{ occurs in }\chi\}\bigr).
\]
Choose disjoint sets of fresh variables
$C=\{c_1,\ldots,c_k\}$ and $C'=\{c'_1,\ldots,c'_k\}$, both disjoint
from $V$, and put $W=V\mathbin{\dot\cup}C\mathbin{\dot\cup}C'$.
Thus $|W|\ge3$. All target assignments have domain exactly $W$.

\subsection{Forcing a product of the active coordinate ranges}

Enumerate $W=\{w_1,\ldots,w_n\}$ and define
\begin{equation}
 \Gamma_{\mathrm{prod}}
 :=\All\bigwedge_{j=2}^n I_{\{w_1,\ldots,w_{j-1}\}}w_j.
 \label{eq:product}
\end{equation}

\begin{lemma}[Product lemma]
\label{lem:product}
If a dependence model $(\mathfrak N,B)$ satisfies
$\Gamma_{\mathrm{prod}}$ and $U_a:=B(a)$ for $a\in W$, then
\[
 B=\prod_{a\in W}U_a.
\]
Here the product denotes the set of all $W$-assignments with their
$a$-value in $U_a$.
\end{lemma}

\begin{proof}
Each assignment in $B$ lies in the indicated Cartesian product by definition.
For the converse, choose arbitrary values $u_j\in U_{w_j}$. There is an
assignment in $B$ taking $u_1$ at $w_1$. If an assignment $s$ has already
been obtained with values $u_1,\ldots,u_{j-1}$, consider any assignment
with $w_j$-value $u_j$. The independence atom true at $s$
supplies an assignment retaining all earlier values
and taking $u_j$ at $w_j$. After finitely many steps, the required
assignment has been obtained.
\end{proof}

\vspace{-1ex}

This result tells us how imposing the truth of $\Gamma_{\mathrm{prod}}$ forces a dependence model to contain the full Cartesian product of the value ranges for all variables as its assignment set. It is an immediate generalization of the three-variable product constraint in the proof of \cite[Theorem~6.10]{BV21}. 

The lemma gives a product of possibly different
ranges $U_a$. A common first-order domain must still be constructed if we are to obtain a standard first-order model.\footnote{This second step was omitted in the cited publication, and we elaborate it here.}

Note that having a common domain means that for all distinnct variables $a,b\in W$ we have that $\forall u \in U_a \exists v \in U_b\, u = v$. This cannot be expressed directly in $\LI$ for lack of an explicit identity predicate, but we can introduce a surrogate, a set of congruence relations which share the essential properties of an identity relation. How this works is shown in the next subsection.

\subsection{A common quotient domain}

For each ordered pair of distinct variables $a,b\in W$, introduce a
fresh binary relation symbol $R_{ab}$ and abbreviate
$e_{ab}:=R_{ab}(a,b)$. Let $\Gamma_{\sim}$ be the conjunction of
\begin{align}
 &\All(e_{ab}\leftrightarrow e_{ba})
 &&(a\ne b),\label{eq:sym}\\
 &\All\Emod_{\{a\}}e_{ab}
 &&(a\ne b),\label{eq:total}\\
 &\All((e_{ab}\land e_{bc})\to e_{ac})
 &&(a,b,c\text{ pairwise distinct}).\label{eq:tri}
\end{align}
There are finitely many such conjuncts.

To analyze these constraints, use the disjoint union
\[
 H:=\coprod_{a\in W}(\{a\}\times U_a).
\]

\begin{remark}[Separation Property] In fact, the systems $\mathsf{LFD}$, $\LI$ and $\LC$ all satisfy the Separation Property. If we make all value ranges for variables disjoint, for instance by adding their variables as `colors' , truth of formulas of the systems remains the same when passing from an assignment in the original model to its obvious companion in the new model. \end{remark}

For different colors $a,b$, put an edge between $(a, u)$ and $(b, v)$
exactly when $R_{ab}^{\mathfrak N}(u, v)$ holds.  Even when raw values in
two ranges coincide, their colored vertices in $H$ are distinct.
Under the product constraint, \eqref{eq:sym} makes this graph undirected,
\eqref{eq:total} gives every vertex a neighbor in every other color,
and \eqref{eq:tri} completes every triangle whose vertices have three
different colors.

Let's analyse what all this means for connected components of the graph.\footnote{These are the obvious candidates for `objects' in our dependence model since they represent equivalence classes of the surrogate for the true identity relation.} First, every component meets every color by the neighbor assumption. Next, consider any two points $(a, u), (b, v)$ with $a, b$ distinct in the same component, i.e., they are connected by a finite sequence of transitions of the above kind.

\begin{lemma}\label{lem:multipartite} Finitely path-connected points are connected by a single $R$-link between them.
\end{lemma}

%\begin{lemma}[Multipartite representation]
%\label{lem:multipartite}
%Let an undirected graph have at least three nonempty color classes,
%with no edges within a color. Suppose every vertex has a neighbor in
%each other color, and every two-edge path with three distinct colors
%has adjacent endpoints. Then every connected component meets every
%color and is complete between distinct colors.
%\end{lemma}

\begin{proof}
Let $p$ and $q$ be vertices of different colors in one component, and choose a
shortest path $v_0=p,v_1,\ldots,v_\ell=q$. If $\ell>1$ and some
$v_i,v_{i+2}$ have different colors, the triangle assumption replaces
these two edges by one, contradicting shortestness. Thus, if $\ell>1$,
the path alternates between two colors. (Just `zip': Looking from the left, we see the color of $p$ at all odd positions, and looking from the right, we see the color of $b$ at all other positions.) Since the endpoints have
different colors, $\ell$ is odd and $\ell\ge3$.

Now choose a third color and a vertex $z$ of that color adjacent to $v_0$. (This exists since we have at least three variables in the languages, and now use the Neighbor assumption.)
Inductively, adjacency of $z$ to $v_i$, together with the edge
$v_i v_{i+1}$, forces adjacency of $z$ to $v_{i+1}$, because these
three vertices have distinct colors. Hence $z$ is adjacent to both
$p$ and $q$, and another triangle completion gives the edge $pq$.
This contradicts shortestness when $\ell>1$. Therefore $p$ and $q$
are adjacent, as required: one single $R$-link connects them.
\end{proof}

\begin{corollary}
\label{cor:quotient}
Suppose a dependence model $(\mathfrak N,B)$ satisfies
$\Gamma_{\mathrm{prod}}\land\Gamma_{\sim}$. Let $Q$ be the set of
connected components described above and let
\[
 \beta_a:U_a\longrightarrow Q,\qquad
 \beta_a(u):=\text{the component of }(a,u).
\]
Every $\beta_a$ is surjective. Moreover, for $a\ne b$ and $s\in B$,
\begin{equation}
 (\mathfrak N,B),s\sat e_{ab}
 \iff \beta_a(s(a))=\beta_b(s(b)).
 \label{eq:quotient-equality}
\end{equation}

\end{corollary}

\begin{proof}
Every component meets every color, proving surjectivity. Next consider the two sides of the above equivalence. From left to right, an edge puts
its endpoints in one component. From right to left, the converse for different colors in the same component 
is \cref{lem:multipartite}.
\end{proof}

\vspace{-3ex}

\subsection{Coherent interpretations of relation symbols}

For each positive-arity source predicate $P$ of arity $r$, introduce a
fresh predicate $\widehat P$ of the same arity and impose
\begin{equation}
 \Gamma_P:=\All\left(
  \bigwedge_{i=1}^r e_{c_i c'_i}
  \ \to\
  \bigl(\widehat P(c_1,\ldots,c_r)
      \leftrightarrow\widehat P(c'_1,\ldots,c'_r)\bigr)
 \right).
 \label{eq:coherence}
\end{equation}
Write $\Gamma_{\mathrm{rel}}:=\bigwedge_P\Gamma_P$, with the conjunction
taken over positive-arity predicates of $\chi$.

\begin{lemma}[Predicate coherence]
\label{lem:coherence}
Under $\Gamma_{\mathrm{prod}}\land\Gamma_{\sim}\land
\Gamma_{\mathrm{rel}}$, the rule
\begin{equation}
 P^{Q}(q_1,\ldots,q_r)
 \iff\widehat P^{\mathfrak N}(u_1,\ldots,u_r),
 \quad u_i\in U_{c_i},\ \beta_{c_i}(u_i)=q_i,
 \label{eq:quotient-predicate}
\end{equation}
defines a well-defined relation on $Q$.
\end{lemma}

\begin{proof}
Representatives exist by surjectivity. Suppose $u_i,v_i\in U_{c_i}$
represent the same $q_i$. Choose $h_i\in U_{c'_i}$ representing $q_i$.
The product lemma supplies an assignment with $c_i=u_i,c'_i=h_i$
for all $i$, and another with $c_i=v_i,c'_i=h_i$ for all $i$.
At both assignments the antecedent of \eqref{eq:coherence} holds by
\eqref{eq:quotient-equality}. Consequently,
\[
 \widehat P^{\mathfrak N}(\bar u)
 \iff\widehat P^{\mathfrak N}(\bar h)
 \iff\widehat P^{\mathfrak N}(\bar v).
\]
Thus the value of \eqref{eq:quotient-predicate} is independent of the
chosen representatives. This defines a single relation for all source
argument patterns, including permutations and repetitions.
\end{proof}

\begin{remark}
The maps $\beta_a$ here need not be injective. Fixed argument positions
alone would therefore not guarantee a well-defined relation on $Q$.
The second set of slots $C'$ and \eqref{eq:coherence} supply exactly
the invariance under changes of representatives that is needed.
\end{remark}

Taking the preceding definitions together, we have now defined a general semantic transformation {\bf F} from dependence models $\mathcal{N}$ satisfying the special condition $\Gamma_{\mathrm{prod}}\land\Gamma_{\sim}\land\Gamma_{\mathrm{rel}}$ to standard models {\bf F}($\mathcal{N}$) for first-order logic with identity.

\vspace{-1ex}

\subsection{Syntactic translation}

As a second part of our intended system connection, we need to connect their logical formulas. We now introduce a syntactic translation taking arbitrary formulas of $\mathsf{FO(=)}$ to formulas of the dependence logic $\LI$. What follows is as in the earlier literature on the logics $\mathsf{CRS}$ and $\mathsf{LFD}$, with a special treatment of atomic formulas.

For a positive-arity atom $P(a_1,\ldots,a_r)$, define
\[
 \Copy^{I}_{\bar a}:=\bigwedge_{i=1}^r e_{a_i c_i}.
\]
All $a_i$ are in $V$ and all $c_i$ are fresh, so each displayed
equality surrogate is defined. Translate first-order formulas by
\begin{align}
 \rho(P(a_1,\ldots,a_r))
   &:=\Emod_V\bigl(\Copy^{I}_{\bar a}\land
                          \widehat P(c_1,\ldots,c_r)\bigr),\notag\\
 \rho(a=b)&:=e_{ab}\quad(a\ne b),\qquad\,\,\,\,\,\,
 \rho(a=a):=\top,\notag\\
 \rho(\neg\psi)&:=\neg\rho(\psi),\qquad \qquad
 \rho(\psi\land\theta):=\rho(\psi)\land\rho(\theta),\notag\\
 \rho(\exists x\,\psi)&:=\Emod_{V\setminus\{x\}}\rho(\psi).
 \label{eq:rho}
\end{align}
For a nullary $P$, set $\rho(P):=\widehat P$. Other connectives and
universal quantification are treated as the usual abbreviations. In the quantifier clause
only the other \emph{source} coordinates are fixed; auxiliary
coordinates may change. Note also that the translation $\rho$ does not use independence atoms, though the special condition $\Gamma_{\mathrm{prod}}\land\Gamma_{\sim}
       \land\Gamma_{\mathrm{rel}}$ does.

%At this stage, we have a syntactic translation $\rho$ as described, plus a special condition  formulated in the language of $\LI$ which singles out the dependence models of interest to us. Note also that independence atoms only occur in this special condition.

%Put \begin{equation} R_I(\chi):=\Gamma_{\mathrm{prod}}\land\Gamma_{\sim} \land\Gamma_{\mathrm{rel}}\land\rho(\chi).\label{eq:ri} \end{equation}

\vspace{-1ex}

\subsection{Analyzing two model constructions} 
We start with a straightforward auxiliary model transformation which motivates the three special conditions that we introduced in the preceding subsections. 

\smallskip

Consider any first-order structure $\mathfrak M$ with object domain $U$. Here is a direct transformation {\bf G} taking such models to dependence models {\bf G}($\mathfrak M$). Take the full team $B = U^W$. Interpret every $R_{ab}$ as the identity relation on $U$, and interpret $\widehat P$ as $P^{\mathfrak M}$.
The product and common-domain constraints hold immediately. The coherence constraint holds because its antecedent makes the two argument tuples literally equal. Note that if $U$ is finite so is the target domain.

\begin{lemma}[Forward connection]
\label{lem:ind-forward}
For every first-order subformula $\psi$ of a given translated formula $\chi$ and for every assignment $s\in B$, we have

\vspace{-1ex}

\[
 \mathfrak M, s|_V\sat\psi
 \iff {\bf G}(\mathfrak M), s\sat\rho(\psi).
\]
\end{lemma}

\begin{proof}
Use induction on first-order formulas. (i) For a relation atom, the copy formula requires $c_i=s(a_i)$, and the
full product supplies these values while fixing $V$. (ii) Equality is
literal identity by construction. (iii) The Boolean cases follow routinely as usual.
(iv) Finally consider the quantifier case. If ${\mathfrak M}, s|_V \sat \exists x \phi$, then there exists an object $u \in U$ such that ${\mathfrak M}, s|_V [x\to u] \sat \phi$. The full product set of assignments permits all choices of the new $x$-value while leaving the other source coordinates fixed. So there is an assignment $s'$ in ${\bf G}(\mathfrak M)$ with $s'(x) = u$ and $s'|_V = s|_V [x\to u]$. By the inductive hypothesis then, ${\bf G}(\mathfrak M), s' \sat \rho(\phi)$. It follows immediately that ${\bf G}(\mathfrak M), s \sat \rho(\exists x\phi)$.
The converse direction is completely analogous. 
\end{proof} 

\smallskip

\begin{remark} Strictly speaking, the preceding lemma is not needed for the main results to follow, and {\bf G} is only auxiliary to our main transformation to follow. Details will be  explained below.

\end{remark}

Next we consider the crucial modal transformation {\bf F}($\mathcal{N}$) introduced earlier taking dependence models $\mathcal{N}$ satisfying our three special constraints $\Gamma_{\mathrm{prod}}\land\Gamma_{\sim}
       \land\Gamma_{\mathrm{rel}}$ to standard first-order models. Notations are as introduced earlier, with components $q$ as the new objects, and the interpretation of predicates $P$ in {\bf F}($\mathcal{N}$) given by the earlier stipulation $P^{Q}(q_1,\ldots,q_r)$ in \cref{lem:coherence}.

\begin{lemma}[Backward connection]
\label{lem:ind-backward}
%Consider a model $\mathcal N = (\mathfrak N,B)$ satisfy the three global constraints in \eqref{eq:ri}. Give $Q$ the predicate interpretations from \cref{lem:coherence}, carrying nullary truth values over unchanged.
For any assignment $s \in B$, define $\bar s:V\to Q$ by
$\bar s(a):=\beta_a(s(a))$. For every subformula $\psi$ of $\chi$,
\begin{equation}
 \textrm{{\bf F}}(\mathcal {N}),\bar s\sat\psi
 \iff \mathcal N,s\sat\rho(\psi).
 \label{eq:ind-truth}
\end{equation}
\end{lemma}

\begin{proof}
We proceed by induction. 

(i) Regular atomic formulas. If $t=_Vs$ satisfies $\Copy^{I}_{\bar a}$, then
\eqref{eq:quotient-equality} yields
\[
 \beta_{c_i}(t(c_i))=\beta_{a_i}(s(a_i))=\bar s(a_i).
\]
Thus any witness for the translated atom supplies representatives of
exactly the source argument tuple. Conversely, choose representatives
in the $c_i$-ranges for this tuple. The product lemma gives a witness
with those auxiliary values and with all source coordinates unchanged.
By predicate coherence, every such choice gives the truth value
specified by $P^{\mathfrak Q}$. This proves the atom equivalence,
including its use underneath negation. 

(ii) The equality cases follow from \eqref{eq:quotient-equality} and reflexivity. 

(iii) The Boolean steps follow routinely  from the induction hypothesis.

(iv) For the existential quantifier step, suppose first that
$\mathfrak Q,\bar s\sat\exists x\,\psi$. Choose a witnessing
$q\in Q$, and choose $u\in U_x$ with $\beta_x(u)=q$.
The full product gives a $t\in B$ with $t(x)=u$ and
$t|_{V\setminus\{x\}}=s|_{V\setminus\{x\}}$.
Then $\bar t=\bar s[x\mapsto q]$, so induction gives
$\mathcal N,t\sat\rho(\psi)$. This witnesses the translated
existential. Conversely, any witness $t$ for that translation agrees
with $s$ on the other source coordinates, hence $\bar t$ agrees with
$\bar s$ there. Its decoded $x$-value witnesses the first-order
existential by induction. \end{proof}

\begin{remark}[A possible simplification] We have reproduced the proof here provided by $\mathsf{ChatGPT}$ with a few minor changes in presentation. However, we believe that the `double copying' using two sets $C$ and $C'$ of auxiliary variables is not necessary for the results. Working with just one copy set $C$ will simplify the above analysis at various places. In particular, the special condition $\Gamma_{\mathrm{rel}}$ for coherent predicate interpretations is no longer needed, and atomic predicates can be defined for components as the new objects just by taking arbitrary representatives. \end{remark}

\vspace{-2ex}

\subsection{The translation from $\mathsf{FO(=)}$ into $\LI$}

Our main result now follows immediately.

\begin{theorem}
\label{thm:fo-ind}
The pair ($\rho$, {\bf F}) is a faithful translation from $\FO(=)$
into $\LI$.
\end{theorem}

\begin{proof}
As explained in Section 1, this follows from (a) the general properties of the transformation {\bf F} listed in  \cref{lem:ind-backward}, plus (b) the fact that  {\bf F} is surjective up to isomorphism. The latter assertion is easy to check.  For any model $\mathfrak{M}$ for $\FO(=)$, first applying the auxiliary transformation {\bf G} and then the transformation {\bf F} yields a model that is isomorphic to $\mathfrak{M}$.\end{proof}

%For the converse, a point satisfying $R_I(\chi)$ lies in a model satisfying its global constraints. By \cref{lem:ind-backward}, its decoded assignment satisfies $\chi$ in $\mathfrak Q$, and therefore $\mathfrak Q\sat\chi$.

An immediate standard consequence of the observations in this proof is that we have found an \textit{effective SAT reduction} from $\FO(=)$ to $\LI$. Here the fact that all variables, predicates, and constraint families are obtained by finite syntactic operations on $\chi$ proves effectiveness.

\vspace{1ex}

\begin{remark}[Finite models] Our two model constructions also turn finite models into finite models. Therefore, we also have an effective embedding of satisfiability on finite models.\footnote{Even so, this observation has only limited value since none of the extended dependence logics considered in this article has the Finite Model Property. The reason is their undecidability, which, in conjunction with the recursive enumerability of their validities through the earlier translations into first-order logic, refutes the FMP.} \end{remark}

%\begin{remark}[The role of the product team]
%When this construction is applied to $S_{\LC}(\varphi)$, the original
%team is represented by the first-order predicate $T$. The new product
%team provides an assignment space for evaluating that first-order
%structure. Its product property places no requirement that the
%original team represented by $T$ was itself a product.
%\end{remark}

\section{Reducing $\FO$ to $\LE$}

In \cite{GP21}, by constructing a conservative satisfiability reduction from the Kahr-Class\footnote{
This is the fragment of $\FO$, usually denoted $[\forall\exists\forall,(\omega,1)]$, which consists of sentences of the form $\forall x_1 \exists y \forall x_2 \varphi(x_1,y,x_2)$ where $\varphi$ is quantifier-free, contains no equality, function symbols or constants, and can only include one binary relation but an unbounded number of unary ones.) }
to $\LE$, it is proved that both the satisfiability and finite satisfiability problems for $\LE$  are undecidable. 

In this section, we present an effective satisfiability reduction from equality-free pure (in the sense of containing no function symbols and constants) first-order logic $\FO$ to $\LE$ discovered by $\mathsf{ChatGPT}$.  Our purpose is not to reestablish the undecidability result in \cite{GP21}. Instead, the reduction is meant to generalize the copying idea employed in \cite{GP21} and uncover the potential of $\LE$ to function as a computing machine.  

\subsection{Source and target of the syntactic mapping}

Every sentence $\alpha$ in $\FO$ can be put into prenex form with its quantifier-free matrix $\theta(\bar{x},\bar{y})$ in negation normal form,\footnote{
In the negation normal form, the negation operator is applied exclusively to atomic predicates and no negations appear over boolean connectives. 
} where the variables in $X = \{x_1,\ldots,x_n\}$ and in $Y = \{y_1,\ldots,y_m\}$ are its universal and existential variables respectively, ordered by their occurrence.
For each $y_i$, let $I_i\subset X$ be the universal variables preceding $y_i$ in the prefix, ordered by their occurrence.

The standard Skolem condition for a sentence $\alpha$ in $\FO$ states that the satisfiability of $\alpha$ is equivalent to the existence of a nonempty relational structure $\B$ which includes functions $f_i: B^{|I_i|}\rightarrow B$ serving as the interpretation of the Skolem functions such that 
\begin{equation}
    \B\models \theta(\bar{a},f_1(\bar{a}_{I_1}),\ldots, f_m(\bar{a}_{I_m}))\qquad \text{for all } \bar{a}\in B^n\enspace .
     \label{skolem}
\end{equation}
So, for the source logic, we only need to consider the sentences of $\FO$ in its prenex form with its matrix $\theta(\bar{x},\bar{y})$ in negation normal form.

As for the target logic, we will follow \cite{GP21} and consider the following fragment $\mathsf{L}(=,D)$ of $\LE$
$$\varphi ::= R\bar{x} \mid \neg R\bar{x}\mid x = y\mid D_X y\mid \Emod_X\varphi\mid \Dmod_X\varphi$$
where the negation operator does not apply to the equality atoms and the atoms $D_Xy$. Formulas in this fragment of $\LE$ are in negation normal form. 

What we will show is that there is a computable syntactic mapping $\mathbf{t}$ such that for any sentence $\alpha$ in $\FO$ in prenex form with its quantifier-free matrix $\theta(\bar{x},\bar{y})$ in negation normal form,  $\alpha$ is satisfiable in $\FO$ if and only if $\mathbf{t}(\alpha)\in \mathsf{L}(=,D)$ is satisfiable in $\LE$. 

\subsection{A preliminary description of the basic idea}

Before presenting the reduction, we give a very rough explanation of the idea behind the reduction. 

Given an arbitrary sentence $\alpha$ of $\FO$ in prenex form with its quantifier-free matrix $\theta(\bar{x},\bar{y})$ in negation normal form, we first need to prove that we can construct a dependence model so that $\mathbf{t}(\alpha)$ is satisfiable given that $\alpha$ is satisfiable. This can be easily done. 

Most of the complexity of the reduction comes from the copying strategy used to prove the other direction. 
Given any dependence model $(\mathfrak{M},A)$\footnote{
Recall that $A\subseteq O^V$ is the set of all admissible assignments.} 
where $\mathbf{t}(\alpha)$ is satisfiable, we try to construct a domain of values $Q\subseteq O$ so that $Q^n\subseteq A(X)$. Here $A(X)$ is the set of all possible value-profiles of the variables in $X$ according to $A$.  The set $Q$ contains all and only the values which interpret the terms in the \emph{Herbrand universe} generated from a constant $\mathbf{c}$ and all the Skolem functions. The syntactic $\mathbf{t}$ should make sure that all the terms in the Herbrand universe can be interpreted as values in some admissible assignments. This can be done by the dependence atoms, representing the functional dependence required by the Skolem functions. Moreover, to make sure $Q^n\subseteq A(X)$, the syntactic mapping $\mathbf{t}$ should make sure that we can copy values between variables so that if the original assignments are admissible then the assignments resulting from copying are still admissible. 

The copying strategy described above has been used in \cite{GP21} to construct the conservative reduction from the Kahr-Class to $\mathsf{L}(=,D)$. Due to the simple structure of the prefix allowed in the Kahr-Class, there is only one Skolem function $f$ to be considered and the terms in the Herbrand universe all have a very simple structure $f^k(\mathbf{c})$. Therefore, once we can make sure that $f^k(c_0)$ appears somewhere, we can first copy it to $x_1$ which should force $y$ to be $f^{k+1}(c_0)$. After that, we can copy $f^{k+1}(c_0)$ from $y$ to the variable whose value is originally $f^k(c_0)$. If $x_1$ is not the final variable where the value $f^{k+1}(c_0)$ should go, then a critical step during our copying process is to store the original value of $x_1$ somewhere else so that we can copy it back later. For this purpose, one more variable in the model in addition to $x_1,$, $x_2$ and $y$ is enough. 

However, for the complete $\FO$ language, the Herbrand universe can be very complex. Here, to make sure that  the denotation of a complex term appears as the value of a particular variable, we need to run through all its sub-terms, a process during which some values of variables need to be stored. All this should be done by copying values among variables. Without a smart management mechanism, as terms becomes more and more complex, we will need more and more variables for storing intermediate values produced in the process, and these do not fit into one dependence model with its fixed finite set of variables. The key to the reduction we will present is a smart use of the data structure of a `stack'. Instead of storing a single value for each variable, the reduction will use some variables to store finite sequences of values and some other variables to store information about the inner structure of these stacks. Moreover, the reduction will make sure that we can simulate the operations on a stack by copying values between variables.  

\subsection{The syntactic mapping}

In the syntactic mapping $\mathbf{t}$, we need fresh variables $z,c,p,q,h,t$, pairwise distinct and disjoint from $X\cup Y$. The following table describes the role of each variable. 
\begin{center}
\begin{tabular}{@{}cl@{}}
\toprule
Variables & Role\\
\midrule
$X$ & Universal data registers\\
$Y$ & Outputs of the Skolem dependency maps\\
$z$ & Accumulator holding a computed term value\\
$c$ & A globally constant seed\\
$p$ & Current stack pointer\\
$q$ & Node being looked up or allocated\\
$h,t$ & Head and tail fields of that node\\
\bottomrule
\end{tabular}
\end{center}

Putting $R=X\cup\{z\}$, we define the following \emph{local} instruction formulas;
the outer $\All$ in the final construction will require every instruction to be
available at every assignment.
\begin{align}
 \Save_j&:=E_{R\cup\{p\}}(h=x_j\land t=p), &&1\leq j\leq n,
 \label{save}\\
 \Write_j&:=E_{(R\setminus\{x_j\})\cup\{p\}}(x_j=z), &&1\leq j\leq n,
 \label{write}\\
 \Read_j&:=E_{(R\setminus\{x_j\})\cup\{p,q\}}(x_j=h), &&1\leq j\leq n,
 \label{read}\\
 \Seed&:=E_{X\cup\{p\}}(z=c),\label{seed}\\
 \Load_i&:=E_{X\cup\{p\}}(z=y_i), &&1\leq i\leq m,
 \label{load}\\
 \Lookup&:=E_{R\cup\{p\}}(q=p),\label{lookup}\\
 \PushPtr&:=E_{R\cup\{q\}}(p=q),\label{pushptr}\\
 \PopPtr&:=E_{R\cup\{q\}}(p=t).\label{popptr}
\end{align}
These instruction formulas describe different operations as indicated by their names. They are fulfilled by making copies of values between variables. For example, when $\Lookup$ holds at an admissible assignment $s$ in a dependence model, by copying the value of $p$ at $s$ into $q$ and keeping the values of the variables in $R\cup \{p\}$ the same, the resulting assignment $s'$ is guaranteed admissible, which means that the operation represented by this formula is executable at $s$. In the next subsection, we will see in detail how $\Save_j$ and $\PushPtr$ together can realize the function of pushing the value of $x_j$ at the current assignment onto the current stack stored in the pointer variable $p$ and how $\Lookup$, $\Read_k$ and $\PopPtr$ together can realize the function of popping the head off the current stack stored in the pointer variable $p$, copying it into the variable $x_j$ and finally storing the tail of the current stack into $p$.

Let $\mathsf{Instr}$ be the conjunction of all above formulas, including all indexed instances.
The target formula is
\begin{equation}
 \boxed{\displaystyle
 \mathbf{t}(\alpha)=
 \All\left(
 \theta(\bar x,\bar y)\land D_{\varnothing}c
 \land D_{\{q\}}h\land D_{\{q\}}t
 \land\bigwedge_{i=1}^{m}D_{I_i}y_i
 \land\mathsf{Instr}
 \right).}
 \label{translation}
\end{equation}

\begin{theorem}
The construction~\eqref{translation} is an effective satisfiability reduction:
\[
 \alpha\text{ is satisfiable}
 \quad\Longleftrightarrow\quad
 \mathbf{t}(\alpha)\text{ is satisfiable in }\mathsf{L}(=,D).
\]
It uses $n+m+6$ variables, $m+3$ dependence atoms, and $3n+m+4$ instruction
formulas. Its size is polynomial in the relational input size.
\end{theorem}

All displayed equalities and dependence atoms occur positively. The only
negated atoms in~\eqref{translation} come from relational literals in
$\theta$. Apart from the original relational vocabulary, the construction
introduces variables but no new predicate or function symbols. It has modal
depth at most two, counting dependence atoms as atomic.

%To prove the ``if" direction of the above theorem, we will prove  that for every sentence $\alpha$ of $\FO$ in prenex form with its quantifier-free matrix $\theta(\bar{x},\bar{y})$ in negation normal form, there is a semantic transformation $\mathbf{T}_\alpha$, such that for all dependence models $\mathcal{M}$ for $\mathsf{L}(=,D)$:
%$$\mathbf{T}_\alpha(\mathcal{M})\models \alpha \quad \text{iff} \quad \mathcal{M}\models \mathbf{t}(\alpha). $$
%Note that this is much weaker than claiming that there is a semantic transformation $\mathbf{T}$ such that for all $\alpha$ and all $\mathcal{M}$ the equivalence holds. 

\subsection{How these axioms implement a persistent stack}

When a sequence $\nu$ is stored in $q$, as a stack written as $a::\pi$, its head $a$ is stored in $h$, and its tail $\pi$ is stored in $t$. This is realized in a dependence model by 
the global dependencies $D_{\{q\}}h,D_{\{q\}}t$, which mean that there are functions $H: A(q)\rightarrow A(h)$ and $L: A(q)\rightarrow A(t)$ respectively satisfying that, for each $s\in A$,
\[
 H(s(q))=s(h),\qquad L(s(q))=s(t).
\]
Recall that when $A$ is the set of admissible assignments in a dependence model, given that the variables in $Z$ are put in an order $Z_1,\ldots, Z_n$, 
$$A(Z) := \{\bar{a}\mid s(Z_i) = \bar{a}_i \text{ for some } s\in A\}.$$

We can think of $p$ as a pointer. The sequence stored in it is the one onto which one can make a push and a pop. The following lemma explains how these operations on a stack can be simulated. 
\Needspace{11\baselineskip}
\begin{lemma}[Matched push and pop]
From any assignment, one can push the current value of $x_j$ (that is, concatenate the value of $x_j$ and the value of $p$, and store the resulting sequence in $p$) while preserving
every variable in $R$. Given that the resulting sequence after the push is now stored in $p$, one can do a matching pop, which restores the head of that sequence to $x_j$ and restores the tail of that sequence
to $p$, while preserving $R\setminus\{x_j\}$.
\end{lemma}

\begin{proof}
Suppose initially $p=\pi$ and $x_j=a$. Use $\Save_j$ to obtain a state with
the same $R,p$ and
\[
 q=\nu,\qquad H(\nu)=a,\qquad L(\nu)=\pi.
\]
Then use $\PushPtr$ to set $p=\nu$, keeping $R,q$ fixed. This is a push.

Later, suppose $\nu$ has been restored to the pointer $p$. Use $\Lookup$ to set
$q=\nu$, retaining $R,p$. The field dependencies recover $h=a,t=\pi$.
Use $\Read_j$ to set $x_j=h=a$ while retaining $q,p$ and every other data
register. Finally $\PopPtr$ sets $p=t=\pi$, retaining $q$ and all of $R$.
Because $q$ stays fixed in the last two steps, its head and tail cannot change.
\end{proof}
Note that there is no need in this construction to recognize an empty stack:
every pop in the proof matches a preceding push.

\subsection{Forward direction: build a dependence model}

\subsubsection{Constructing the domain of values}
Assume $\alpha$ has a model, and fix one of its Skolem expansions $\B$ satisfying
\eqref{skolem}. Choose a seed $b_0\in B$. Let $S=\Lists(B)$
be the set of all finite lists of elements of $B$, including
the empty list $\varepsilon$. Take the enlarged universe $M=B\mathbin{\dot\cup}S$\footnote{
Note that a sequence of length one is $\langle a \rangle$, which is thus different from the value $a$ itself. 
} 
as the domain of values in our targeted dependence model $G(\B)$. Interpret every original relation on tuples from $B$ exactly as in $\B$,
and arbitrarily on other tuples. Putting the enlarged universe and the interpretations of the original relations together, we get the first-order structure $\mathfrak{M}$ in our target dependence model $G(\B)$. We still need to construct the set of admissible assignments $A$. 

\subsubsection{Constructing the set of admissible assignments}
Define two list operations $\Hd$ and $\Tl$ by
\[
 \Hd(a::\pi)=a,\qquad\Tl(a::\pi)=\pi,
 \qquad \Hd(\varepsilon)=b_0,\quad\Tl(\varepsilon)=\varepsilon.
\]
These are semantic operations used to construct the dependence model; they are not
symbols of the target formula.

Let the set of admissible assignments $A$ contain all assignments $s$ that are obtained by
choosing values for variables independently as follows:
\[
 s(X)\in B^n,\qquad s(z)\in B,\qquad s(p),s(q)\in S,
\]
and then setting
\begin{align*}
 s(c)&=b_0,\\
 s(y_i)&=f_i(s(I_i)),\\
 s(h)&=\Hd(s(q)),\qquad s(t)=\Tl(s(q)).
\end{align*}
The matrix holds at every assignment by~\eqref{skolem}, and all dependence
requirements hold by definition.

\subsubsection{Verifying all instruction formulas hold globally}
Each instruction is available in this team:
\begin{itemize}
 \item $\Save_j$ chooses $q=s(x_j)::s(p)$, preserving $R,p$.
 \item $\Write_j$ and $\Read_j$ set $x_j$ to the indicated data value,
 preserving their specified registers and recomputing any dependent $y_i$.
 \item $\Seed$ and $\Load_i$ set $z$ to the indicated value in $B$.
 \item $\Lookup$ chooses $q=p$; $\PushPtr$ chooses $p=q$; $\PopPtr$
 chooses $p=\Tl(q)$.
\end{itemize}
All choices stay within the indicated independent coordinate ranges. 

Therefore, $G(\B)=(\mathfrak{M},A)\models\mathbf{t}(\alpha)$.

\subsection{Reverse direction: evaluate all ground Skolem terms}

Assume $\mathbf{t}(\alpha)$ is satisfied in a dependence model $\mathcal{M} = (\mathfrak{M},A)$. By the global
dependency on $c$, its value is a fixed element $c_0\in M$. Each 
global dependency specified in $\mathbf{t}(\alpha)$ defines a partial map
\[
 f_i^A:A(I_i)\longrightarrow M,
 \qquad f_i^A(s(I_i))=s(y_i)
\]
corresponding to a Skolem function $f_i$ of $\alpha$.
Choose arbitrary total extensions $\widehat f_i:M^{r_i}\to M$ of these maps;
for example, use $c_0$ on unobserved tuples. The extensions are a device for defining term values before proving that all
relevant tuples are actually observed in $A$.

Form ground terms using a formal
constant $\mathbf c$ and these function symbols, and evaluate each ground term $\tau$ with
$\mathbf c$ interpreted by $c_0$, denoted by $\llbracket\tau\rrbracket$.

\begin{lemma}[Term evaluation with register preservation]\label{lem:evaluateTerm}
For every ground term $\tau$ and every starting assignment $s\in A$,  a process of evaluating a finite sequence of instruction formulas ends at an assignment $u\in A$ with
\begin{equation}
 u(X)=s(X),\qquad u(p)=s(p),\qquad
 u(z)=\llbracket\tau\rrbracket.
 \label{evalinvariant}
\end{equation}
Every function application encountered in this evaluation uses an argument
tuple in the corresponding observed domain $A(I_i)$.
\end{lemma}

\begin{proof}
Proceed by structural induction on $\tau$. For $\tau=\mathbf c$, use $\Seed$. It preserves $X,p$.

For $\tau=f_i(\tau_1,\ldots,\tau_{r_i})$, let the ordered dependency coordinates
be $I_i=(x_{j_1},\ldots,x_{j_{r_i}})$. Execute the following finite procedure $\Eval$ on $\tau$:
\begin{enumerate}
 \item Push the current $x_1,\ldots,x_n$, in that order, using
 $\Save_j$ followed by $\PushPtr$.
 \item For each $\ell=1,\ldots,r_i$, call $\Eval(\tau_\ell)$, which  evaluates $\tau_\ell$ and stores its interpretation into $z$, and use $\Write_{j_\ell}$ to store it in $x_{j_\ell}$.
 (Each recursive evaluation preserves all currently installed $X$-values and
 the current pointer.)
 \item Use $\Load_i$ to copy $y_i$ into $z$.
 \item Pop $x_n,\ldots,x_1$, in reverse order, using $\Lookup$,
 $\Read_j$, and $\PopPtr$ for each saved register.
\end{enumerate}
An example after this proof illustrates how the above algorithm works. 
After step 2 of the algorithm there is an actual team assignment with
\[
 x_{j_\ell}=\llbracket\tau_\ell\rrbracket
 \quad(1\leq\ell\leq r_i).
\]
Thus this argument tuple lies in $A(I_i)$. The value of $y_i$ there is
\[
 f_i^A(\llbracket\tau_1\rrbracket,\ldots,
       \llbracket\tau_{r_i}\rrbracket)
 %=\widehat f_i(\llbracket\tau_1\rrbracket,\ldots,
       \llbracket\tau_{r_i}\rrbracket).
\]
Step 3 preserves $X$, so the dependency $D_{I_i}y_i$ keeps this value stable
while it is copied into $z$. Step 4 restores the original $X,p$ by the stack
lemma and preserves $z$. This proves~\eqref{evalinvariant}.

For a nullary $f_i$, the argument loop is empty and the same reasoning applies.
No value assigned arbitrarily on an unobserved tuple is used in any procedure $\Eval(\tau)$.
\end{proof}

\begin{example}
For
\[
 \alpha=\forall x_1\forall x_2\exists y\forall x_3\,
 \varphi(x_1,x_2,y,x_3),
\]
take $X=\{x_1,x_2,x_3\}$, $Y=\{y\}$, and $I_y=\{x_1,x_2\}$. The complete
target variable set is
\[
 \{x_1,x_2,x_3,y,z,c,p,q,h,t\}.
\]
Use~\eqref{translation} with matrix $\varphi$ and the corresponding instruction
instances. There are four dependence requirements and fourteen instruction
formulas, all inside one global modality along with the original matrix.

For example, evaluating a binary term $f(\rho,\sigma)$ proceeds as follows:
\[
\begin{array}{l}
 \text{push }x_1,x_2,x_3;\\
 \Eval(\rho);\quad x_1:=z;\\
 \Eval(\sigma);\quad x_2:=z;\\
 z:=y;\\
 \text{pop }x_3,x_2,x_1.
\end{array}
\]
The second recursive evaluation $\Eval(\sigma)$ preserves the first argument already stored 
in $x_1$ after the first recursive evaluation $\Eval(\rho)$. The final pops preserve the computed result in $z$. Since the function depends only on
$\{x_1,x_2\}$, $x_3$ never affects the function value.

\end{example}

Now we can construct the first-order structure where $\alpha$ is satisfiable. Define the following generated domain:

\vspace{-3ex}

\[
 Q=\{\llbracket\tau\rrbracket:
       \tau\text{ is a ground term over }\mathbf c,f_1,\ldots,f_m\}.
\]
It is nonempty and closed under every $\widehat f_i$ by term formation.
We claim
\begin{equation}
 Q^n\subseteq A(X).
 \label{cube}
\end{equation}
Given any $a_1,\ldots,a_n\in Q$, choose terms $\tau_j$ interpreted by them respectively. From
any starting assignment $s$, by Lemma \ref{lem:evaluateTerm}, we can evaluate $\tau_1$ into $z$ and apply $\Write_1$;
then evaluate $\tau_2$ and apply $\Write_2$; continue in this way. Each
evaluation in this procedure preserves all $X$-values, and each write preserves the other
coordinates. The final assignment $u$ has $u(X)=(a_1,\ldots,a_n)$, proving
\eqref{cube}. If $n=0$, this is simply the nonemptiness of $A$.

Because every $I_i$ is a subset of $X$, the cube implies that each
$Q^{r_i}$ is observed in $A(I_i)$. Consequently
\[
 f_i^A\upharpoonright Q^{r_i}
 =\widehat f_i\upharpoonright Q^{r_i}:Q^{r_i}\to Q.
\]

\vspace{0.1ex}

For any universal tuple $\bar a\in Q^n$, choose a witnessing team assignment.
The dependency conditions force its $Y$-values to be precisely these Skolem
values, all in $Q$. The global matrix then gives
\[
 \M\upharpoonright A\models\theta\bigl(\bar a,
 \widehat f_1(\bar a_{I_1}),\ldots,
 \widehat f_m(\bar a_{I_m})\bigr)\vspace{0.3ex}
\]

for all $\bar{a}\in Q^X$ where $\mathfrak{M}$ is the first-order relational structure in the dependence model we start with. 
%Restrict the original relational structure to $A$. Quantifier-free relational formulas, including negated relational atoms, are preserved on tuples from an induced substructure. 
Thus $\mathbf{T}(\mathcal{M}) = \M\upharpoonright A$, with these Skolem functions,
satisfies~\eqref{skolem}, and hence satisfies $\alpha$. This completes the
reverse implication and the theorem.

\subsection{Final remarks}

\subsubsection{Finite satisfiability is not preserved}

In the proof of the forward direction, the list domain is
generally infinite, even when $B$ is finite. This is a substantive limitation, not just a choice of presentation. If $n\geq1$ and the extracted domain $A$
contains two elements, the axioms force infinitely many pointer values.
Indeed, evaluation and writing can place any $a\in A$ into $x_1$ while retaining
an arbitrary current pointer $\pi$. A save and push then produce a pointer
$\nu$ with $H(\nu)=a,L(\nu)=\pi$. Lookup and pointer copying give
$A(p)=A(q)$, so the following inequality must hold:
\[
 |A(p)|\geq |A|\,|A(p)|,
\]
which is impossible when $|A|\geq 2$.

For instance,
\[
 \forall x\exists y_0\exists y_1\,
       \bigl(Q(x)\land P(y_0)\land\neg P(y_1)\bigr)\vspace{0.3ex}
    \]
has a two-element model, but the formula it maps to has no finite dependence
model. The generated domain contains $f_0(c_0)$ and $f_1(c_0)$, satisfying $P$ and $\neg P$, respectively, so these values are distinct and the counting argument applies. 

The theorem is therefore a direct \emph{SAT} reduction, rather than the conservative
finite-SAT reduction proved for the Kahr class in Theorem~10 of~\cite{GP21}.

\subsubsection{Less than a translation but more than a satisfiability reduction}

The syntactical mapping $\mathbf{t}$ and the model transformation $\mathbf{T}$ in our proof do not constitute a translation as defined in the Introduction. For, the latter requires the transformation to be defined  independently from the formula $\alpha$. However, the transformation in our proof depends on the syntactical structure of $\alpha$, because the domain $Q$ of the relational structure is constructed using the Skolem functions extracted from $\alpha$. 

In our syntactical mapping $\mathbf{t}$ of $\alpha$, we cannot single out a $\Gamma$ part  that is independent from $\alpha$, as in the translation in the preceding section. However, the embedding given in this section conveys more information than a pure satisfiability reduction. It not only reveals the connection between dependence atoms in $\LE$ and Skolem functions in $\FO$,  but also shows how variables can be viewed as registers in a computing machine and how equality can be used to write and read between registers. Moreover, using a stack, it shows how complex data structures can be simulated. These components together deepen our understanding of the strength of $\LE$.

\section{Translating $\LE$ into $\LC$}
\label{sec:eq-cond}

In this section we provide a faithful translation from the logic $\LE$
of dependence and equality into the logic $\LC$ of conditional dependence.
As in Section~2, we connect the syntactic translation with two model
constructions.

\subsection{The main result}

\begin{theorem}
There is a faithful translation from $\LE$ into $\LC$.
Only atomic conditions are needed in the conditional dependence atoms.
\end{theorem}

Let $\varphi$ be an $\LE$-formula. Let $V$ contain all variable names
occurring in $\varphi$, adding an unused variable if needed. Set
\[
 k:=\max\bigl(\{1\}\cup
       \{\operatorname{arity}(P):P\text{ occurs in }\varphi\}\bigr).
\]
Choose fresh $C=\{c_1,\ldots,c_k\}$ and put
$W=V\mathbin{\dot\cup}C$.
Fix an enumeration $\bar w$ of $W$. Introduce fresh $|W|$-ary predicates
$O,\Delta$ and write
\[
 o:=O(\bar w),\qquad d:=\Delta(\bar w).
\]
The formula $o$ selects assignments representing the original model,
while $d$ selects assignments that will serve as objects in the model
constructed by {\bf F}. Since both predicates use all variables in $W$,
they can select arbitrary subsets of the team. The two subsets may overlap.

\subsection{A common domain}

The first step is to obtain a common domain from the value
ranges of the variables. We use the assignments satisfying $d$ as the
objects of this domain. The following constraint says that every value
occurs at such an assignment, and that each such assignment is determined
by its value at any one variable. Define
\begin{align}
 \Gamma_\Delta
 &:=\All\left(
       \bigwedge_{a\in W}\Emod_{\{a\}}d
       \ \land\
       \left(d\to\bigwedge_{a,b\in W}D_{\{a\}}^d b\right)
       \right),\label{eq:gamma-diagonal}\\
 \Eq(a,b)&:=\Emod_{\{a,b\}}d.\label{eq:diag-eq}
\end{align}

\begin{lemma}[Common domain]
\label{lem:diagonal}
Suppose $\mathcal N=(\mathfrak N,B)\sat\Gamma_\Delta$ and set
$Q:=\{q\in B:\mathcal N,q\sat d\}$. For each $a\in W$ and
$u\in B(a)$ there is a unique $q\in Q$ with $q(a)=u$. Consequently
\[
 \beta_a:B(a)\longrightarrow Q,\qquad
 \beta_a(u):=\text{that unique }q,
\]
is a bijection with inverse $q\mapsto q(a)$. Moreover,
\begin{equation}
 \mathcal N,s\sat\Eq(a,b)
 \iff \beta_a(s(a))=\beta_b(s(b)).
 \label{eq:diag-decoded}
\end{equation}
\end{lemma}

\begin{proof}
Choose $s\in B$ with $s(a)=u$. The conjunct $\Emod_{\{a\}}d$
supplies existence. If $q,q'\in Q$ have the same $a$-value, every
atom $D_{\{a\}}^d b$ holds at $q$ and applies to $q'$. Hence they
agree at all $b\in W$ and are identical. This proves uniqueness.
The inverse formula follows immediately and proves bijectivity.

A witness for $\Eq(a,b)$ is simultaneously the unique representative
of $s(a)$ in coordinate $a$ and of $s(b)$ in coordinate $b$.
Conversely, when these representatives are equal, their common value
is the required witness. This proves \eqref{eq:diag-decoded}.
\end{proof}

\subsection{Interpreting relation symbols}

As in Section~2, we evaluate every occurrence of a relation symbol
using the same auxiliary variables. For each positive-arity predicate
$P$ of arity $r$ occurring in $\varphi$, introduce a fresh predicate
$\widehat P$ of the same arity. For every such atom
$P(a_1,\ldots,a_r)$ occurring anywhere in $\varphi$, put
\begin{align}
 \Copy^C_{\bar a}
   &:=\bigwedge_{i=1}^r\Eq(a_i,c_i),\notag\\
 \Gamma_{\mathrm{copy}}
   &:=\bigwedge_{P(\bar a)\text{ occurs in }\varphi}
       \All\bigl(o\to\Emod_V\Copy^C_{\bar a}\bigr).
 \label{eq:diag-copy}
\end{align}
An empty conjunction is $\top$. At each assignment satisfying $o$,
this constraint lets us choose the required values for $c_1,\ldots,c_r$
while keeping all variables in $V$ fixed. 

We now define a transformation {\bf F} taking dependence models satisfying
these constraints to dependence models for $\LE$. Consider a model
$\mathcal N=(\mathfrak N,B)$ satisfying
$\Gamma_\Delta\land\Gamma_{\mathrm{copy}}$, and assume that some assignment
satisfies $o$. Take $Q$ from \cref{lem:diagonal} as the object domain.
For $s\in B$, define $\bar s:V\to Q$ by
\[
 \bar s(a):=\beta_a(s(a))\quad(a\in V),\qquad
 A_{\mathcal N}:=\{\bar s:s\in B,\ \mathcal N,s\sat o\}.
\]
This team is nonempty by our assumption that there is an assignment satisfies $o$. Define a relational structure
$\mathfrak Q$ on $Q$ by
\begin{equation}
 P^{\mathfrak Q}(q_1,\ldots,q_r)
 \iff\widehat P^{\mathfrak N}(q_1(c_1),\ldots,q_r(c_r)),
 \label{eq:diagonal-predicate}
\end{equation}
and carry over nullary truth values. Each $q_i\in Q$ is itself a
$W$-assignment, so the expression $q_i(c_i)$ is defined. Put
\[
 {\bf F}(\mathcal N):=(\mathfrak Q,A_{\mathcal N}).
\]
This completes the definition of the semantic transformation {\bf F}.
The corresponding source assignment is $\bar s$, for each $s$ satisfying $o$.

\subsection{Syntactic translation}

As a second part of our intended system connection, we now introduce a
syntactic translation from $\LE$ to $\LC$. Equality is interpreted on the
common domain $Q$, and relation atoms use the auxiliary variables just
introduced. Translate by
\begin{align}
 \tau(P(a_1,\ldots,a_r))
   &:=\Emod_V\bigl(\Copy^C_{\bar a}\land
                             \widehat P(c_1,\ldots,c_r)\bigr),\notag\\
 \tau(a=b)&:=\Eq(a,b),\notag\\
 \tau(\neg\psi)&:=\neg\tau(\psi),\qquad
 \tau(\psi\land\theta):=\tau(\psi)\land\tau(\theta),\notag\\
 \tau(\Dmod_X\psi)&:=\Dmod_X(o\to\tau(\psi)),\notag\\
 \tau(D_Xy)&:=D_X^o y.
 \label{eq:tau}
\end{align}
Nullary atoms are copied to fresh nullary predicates without an
auxiliary witness. The Boolean connectives not displayed here are
abbreviations.

Let $\Gamma_C:=\Gamma_\Delta\land\Gamma_{\mathrm{copy}}\land o$. The condition $\Gamma_C$ restricts the target models and assignments
used in the translation.

Next we show how this syntactic translation interacts with our relevant model constructions.

\subsection{Two model constructions}

We start with a straightforward transformation which motivates the two
special conditions introduced above. Consider any dependence model
$\mathcal M=(\mathfrak M,A)$ for the predicates occurring in $\varphi$,
with object domain $U$ and assignments on $V$. Here is a direct
transformation {\bf G} taking such models to dependence models for $\LC$.
Write ${\bf G}(\mathcal M)=(\mathfrak N,B)$. Use the same domain $U$ and define
\begin{align*}
 B_o&:=\{s:W\to U:s|_V\in A\},\\
 B_d&:=\{d_u:u\in U\},\qquad
                         d_u(a):=u\quad(a\in W),\\
 B&:=B_o\cup B_d.
\end{align*}
Interpret $O$ and $\Delta$ so that $o$ holds exactly at the assignments
in $B_o$ and $d$ exactly at those in $B_d$. For each source predicate $P$,
put $\widehat P^{\mathfrak N}:=P^{\mathfrak M}$.

\begin{lemma}[Forward connection]
\label{lem:cond-forward}
The model ${\bf G}(\mathcal M)$ satisfies both global constraints. Moreover, for every $s\in B$,
\[
 {\bf G}(\mathcal M),s\sat\Eq(a,b)\iff s(a)=s(b).
\]
For every subformula $\psi$ of $\varphi$ and every $s\in B_o$,
\begin{equation}
 \mathcal M,s|_V\sat\psi
 \iff {\bf G}(\mathcal M),s\sat\tau(\psi).
 \label{eq:cond-forward-truth}
\end{equation}
\end{lemma}

\begin{proof}
For each $s,a$, the assignment $d_{s(a)}$ witnesses
$\Emod_{\{a\}}d$. Two constant assignments agreeing at one coordinate
agree everywhere, which proves all required dependences at assignments
satisfying $d$. Thus $\Gamma_\Delta$ holds. A constant assignment agrees with $s$ at
both $a,b$ exactly when $s(a)=s(b)$, proving the equality assertion.

At an assignment $s\in B_o$, fix $s|_V$ and choose $c_i=s(a_i)$ for
any source atom under consideration. Complete unused auxiliary
coordinates arbitrarily in the nonempty $U$. (Since we are considering the subformulas of $\psi$.) The resulting assignment
belongs to $B_o$ and satisfies the corresponding copy formula.
Hence $\Gamma_{\mathrm{copy}}$ holds.

We prove the truth equivalence by induction on formulas.

(i) Regular atomic formulas. Every assignment agreeing with $s$ on $V$
and satisfying the copy formula has $c_i=s(a_i)$, and such assignments
always exist. Since
$\widehat P$ is interpreted as $P^{\mathfrak M}$, this gives the required
atom equivalence.

(ii) The equality case follows from the equality assertion above.

(iii) The Boolean steps follow routinely from the induction hypothesis.

(iv) For $\Dmod_X\psi$, the assignments satisfying $o$ and agreeing
with $s$ on $X\subseteq V$ restrict to exactly the assignments in $A$
agreeing with $s|_V$ on $X$. Apply the induction hypothesis to obtain
the modal equivalence.

(v) The same correspondence, together with unchanged $y$-values,
identifies $D_X^o y$ with the original dependence atom. This completes
the induction.
\end{proof}

\begin{remark}[Nonempty variable sets]
The non-emptiness of $W$ is used to make sure $Q$ is non-empty. For any $a\in W$, the conjunct
$\All\Emod_{\{a\}}d$ ensures that $Q$ is nonempty, while $d_u(a)=u$
makes $u\mapsto d_u$ injective. If $W=\varnothing$, $\Gamma_\Delta$
is vacuous, so {\bf F} may yield an empty object domain.
\end{remark}

Next we consider the transformation {\bf F} introduced earlier, taking
models $\mathcal N=(\mathfrak N,B)$ satisfying the two global constraints
and having an assignment satisfying $o$ to models for $\LE$.
The domain $Q$, the maps $\beta_a$, and the assignments $\bar s$ are
as defined above.

\begin{lemma}[Backward connection]
\label{lem:cond-backward}
For every $s\in B$ with $\mathcal N,s\sat o$ and  subformula
$\psi$ of $\varphi$,
\begin{equation}
 {\bf F}(\mathcal N),\bar s\sat\psi
 \iff\mathcal N,s\sat\tau(\psi).
 \label{eq:cond-backward-truth}
\end{equation}
\end{lemma}

\begin{proof}
First, injectivity of each $\beta_a$ gives, for assignments $s,t$
satisfying $o$ and for $X\subseteq V$,
\begin{equation}
 s=_Xt\iff\bar s=_X\bar t.
 \label{eq:injective-agreement}
\end{equation}
We now prove \eqref{eq:cond-backward-truth} by induction.

(i) Regular atomic formulas. Any $t\in B$, $t=_Vs$ satisfying $\Copy^C_{\bar a}$ has
\[
 \beta_{c_i}(t(c_i))
   =\beta_{a_i}(s(a_i))=\bar s(a_i).
\]
The inverse of $\beta_{c_i}$ is evaluation at $c_i$, so every such
assignment has the same required values
\begin{equation}
 t(c_i)=\bar s(a_i)(c_i).
 \label{eq:unique-slots}
\end{equation}
Now we prove the left-to-right direction, if $\mathcal N,s\sat\tau(P(\bar a))$, the source atom
holds at $\bar s$ by \eqref{eq:diagonal-predicate}. Conversely, when the source
atom is true, $\Gamma_{\mathrm{copy}}$ supplies an assignment satisfying
the copy formula, and \eqref{eq:unique-slots} forces the predicate to be
true there as well. This proves the atom equivalence, also for atoms
occurring underneath negation.

(ii) The equality case is precisely \eqref{eq:diag-decoded}.

(iii) The Boolean steps follow routinely from the induction hypothesis.

(iv) For $\Dmod_X\psi$, every assignment satisfying $o$ and agreeing
with $s$ on $X$ gives an assignment in ${\bf F}(\mathcal N)$ agreeing
with $\bar s$ on $X$. Conversely, every such assignment in
${\bf F}(\mathcal N)$ is $\bar t$ for some $t$ satisfying $o$;
\eqref{eq:injective-agreement} ensures that $t$ agrees with $s$ on $X$.
Assignments not satisfying $o$ satisfy $o\to\tau(\psi)$ vacuously.
Apply the induction hypothesis to obtain the modal equivalence.

(v) Finally, $D_X^o y$ quantifies exactly over assignments satisfying
$o$ and agreeing with $s$ on $X$. Their corresponding assignments in
${\bf F}(\mathcal N)$ are exactly those quantified over by $D_Xy$
at $\bar s$, and injectivity of $\beta_y$ identifies equality of their
$y$-values. This proves the dependence case and completes the induction.
\end{proof}

\begin{remark}[On auxiliary variables $C$]

In the translations of Sections 2 and 4, the role of the auxiliary variables $C$ is best
seen in the \emph{backward construction} {\bf F}. Starting with a target
dependence model $\mathcal N=(\mathfrak N,B)$ satisfying the respective
global constraints, we reconstruct a source model on the common domain
$Q$. In each construction, $\beta_a$ maps values in the $a$-coordinate
to objects of $Q$, and the decoded source assignment is
$\bar s(a)=\beta_a(s(a))$. The interpretation of each source predicate
must therefore depend only on the tuple of decoded objects.

For example, in the backward construction for the present translation
from $\LE$ to $\LC$, \eqref{eq:diag-decoded} says that $\Eq(x,y)$
expresses $\bar s(x)=\bar s(y)$, which need not imply $s(x)=s(y)$.
Thus the naive translation
$\Eq(x,y)\land\widehat P(x)\land\neg\widehat P(y)$ may be satisfiable even
though the source formula $x=y\land P(x)\land\neg P(y)$ is
unsatisfiable.

The variables $C$ provide fixed argument positions for interpreting
predicates on $Q$. In the present backward construction, each
$\beta_{c_i}$ is a bijection by \cref{lem:diagonal}, and the unique
representative of $q_i\in Q$ in the $c_i$-coordinate is $q_i(c_i)$.
This is exactly the value used to define $P^{\mathfrak Q}$ in
\eqref{eq:diagonal-predicate}. At an assignment $s$ satisfying $o$,
$\Gamma_{\mathrm{copy}}$ supplies a $t=_Vs$ satisfying
$\Copy^C_{\bar a}$. By \eqref{eq:unique-slots}, every such $t$ has
\(
 t(c_i)=\bar s(a_i)(c_i)\quad(1\le i\le r).
\)
Consequently, the witness in the atomic clause of \eqref{eq:tau}
tests $\widehat P$ on exactly the tuple used to interpret
$P^{\mathfrak Q}(\bar s(a_1),\ldots,\bar s(a_r))$.
This is the atomic step of \cref{lem:cond-backward}.

In particular, consider the example above at an assignment $s$
satisfying $o\land\Eq(x,y)$, and put $q=\bar s(x)=\bar s(y)$.
By \eqref{eq:unique-slots}, copying either $x$ or $y$ into $c_1$
yields the same value $q(c_1)$.
Since $\Gamma_{\mathrm{copy}}$ guarantees the required copying
witnesses in both cases, the translations of $P(x)$ and $P(y)$
both test whether $q(c_1)\in\widehat P^{\mathfrak N}$.
They therefore have the same truth value, ruling out truth at $s$ for the
translation of the formula $x=y\land P(x)\land\neg P(y)$.
\end{remark}

\subsection{The translation from $\LE$ into $\LC$}

Our main result now follows immediately.

\begin{theorem}\label{thm:faithLEtran}
The pair \((\tau,\mathbf F)\) is a faithful translation from $\LE$ into $\LC$.
Only atomic conditions are needed in the conditional dependence atoms.
\end{theorem}

\begin{proof}
Apply the construction to the conjunction of the premises and the conclusion, as in Remark 1.4. \cref{lem:cond-backward} gives the required truth equivalence. For every source model \(\mathcal M\), the map \(u\mapsto d_u\) induces an isomorphism \(\mathcal M\cong F(G(\mathcal M))\), and every source assignment has an extension in \(B_o\). Thus \(\bf F\) is surjective up to isomorphism on pointed models, and faithfulness of the translation  follows as in Section 1.
\end{proof}

As a corollary of \cref{thm:faithLEtran}, the map \(\varphi\mapsto\Gamma_C\land\tau(\varphi)\), with \(\Gamma_C\) constructed for \(\varphi\), is an effective SAT reduction from \(\LE\) to \(\mathsf{LFD}+\mathsf{Con}\).

\section{Discussion and general directions}

In this final section we  add some observations to the main results reported here. 

\smallskip

{\bf Richer dependence logics.}  This paper has focused on the three extended dependence logics introduced in \cite{BV21}. Some natural extensions of these all still translate naturally into first-order logic with identity, such as a combination of $\LC$ and $\LI$ which also allows conditional independence atoms. We believe that the proof methods presented in this Note still apply to drawing a system like this into our circle of reduction and translation results.

Another natural extension, still first-order translatable like the systems we considered here, comes from the standard  literature in modal logic. Note that conditional dependence atoms $D^{\varphi}_Xy$  can also be seen as adding a new modality to $\mathsf{LFD}$. Their defining local truth condition  $\forall t:\, (s =_X t \land t \models \varphi) \rightarrow s =_y t$ as stated earlier is logically equivalent to 

\vspace{-1ex}

$$\forall t:\, (s =_X t \land \neg s =_y t) \rightarrow  t \models \neg\varphi$$. 

\vspace{-3ex}

But this is definable in Boolean modal logic. This is a system with modalities for any combination of accessibility relations using Boolean set operations, in our case ($\mathsf{S5BML}$): starting from basic equivalence relations, which was axiomatized in \cite{SCH26}. It follows from our results that $\mathsf{S5BML}$ is undecidable.

\smallskip

{\bf New insights from our SAT reductions and translations.}  The case studies offered here  enrich the original understanding of dependence logics in \cite{BV21}. For instance, in the translation from $\LE$ into $\LC$, the surprising power of conditional dependence shows in its ability to produce assignments giving every variable the same object value, and then using existential dependence modalities to encode identity of variables representing objects. Also, the translation from $\mathsf{FO(=)}$ into $\LI$ introduces an innovative use of a family of binary relations mimicking identity of objects in value ranges for variables, and then allowing an unusual style of quotienting for creating uniform objects. And even more strikingly, the SAT reduction from $\mathsf{FO(=)}$ into $\LE$ showed how, with data stacks as values for auxiliary variables, dependence models can act as state-based computing devices for term values in the style of push-down store automata. 

%In principle, a translation of the sort in this note is just a correlation of semantic evaluation of formulas in correlated models for two logics, and in our case, one that only works when the models on one side satisfy some special conditions. 
\smallskip

{\bf Logical system connections.} Our results in this Note are a mixture of SAT reductions and faithful translations between extended dependence logics. As noted in Section 1, the latter can transfer more system properties than just satisfiability. For several of the logical systems considered here, the existence of a faithful translation is an open problem. 

But perhaps the more interesting general methodological issue is this. The results in this Note are either effective SAT reductions or faithful translations. But the proofs for these results suggest a richer spectrum of connections between logical systems than just these two. For instance,  the standard translations of our  dependence logics into $\mathsf{FO(=)}$ get close to direct definability on the same models, as happens in the standard translation for the basic modal logic, be it modulo one natural expansion with an additional `team predicate'. Next, several translations that we found use a `Chu schema'  which tightly correlates semantic evaluation of formulas in the source language with semantic evaluation of their translations in suitable correlated models:

\vspace{-1ex}

\begin{quote}

 {\bf T}($\mathfrak{M})\models \varphi$ \quad iff \quad $\mathfrak{M}\models$ {\bf t}($\varphi$). 

\end{quote}

\vspace{-1ex}

But also among conservative SAT reductions, there is variety. For instance, the tiling reduction in Appendix A only correlates satsifiability in $\mathsf{FO}(=)$ and in $\LC$ in the weak sense that tiling problems correlate with satisfiability of first-order formulas via computational coding. But our SAT reduction from $\mathsf{FO}$ to $\LE$ stayed much closer to the syntax of the two logics, directly correlating first-order formulas in prenex form or Skolem form with dependence formulas.\footnote{In the realm of Skolem forms, Herbrand models, and resolution theorem proving, satisfiability problems may be a better focus anyway than the usual logical consequence problems highlighted in translations. } 

We merely raise the issue of a more fine-grained hierarchy of logical system connections here, without going into detail, for instance, on how close `logical SAT reductions can get to faithful translations, or faithful translations to direct definitions.\footnote{ For an attempt at classifying styles of translating between logical systems, see \cite{RF10}.} 

%The faithful translations into $\LC, \LE$ and $\LI$ that were presented here established close connections between logical systems. And the reverse connections were even tighter, asWe have refuted real direct definability in Appendix B, but one could also think about weaker correlations of logical systems than the ones used here. One prominent example, of course, are SAT reductions which can work with any correlation between undecidable problems and encoding the relevant computational structure in a logic.
%
\medskip

{\bf The AI experience.} We started this project  as a test case for seeing first-hand what AI can do for logical research. The impetus for this were results announced in September of this year on solving open problems in logic by AI techniques.\footnote{For two lively interviews with the people doing this, see the wide-ranging  survey book \cite{BHIL26}.} We started by asking $\mathsf{ChatGPT}$ for SAT reductions between our dependence logics, with a view to establishing their decidability or undecidability. As the interaction continued, we found other types of result, and further questions emerged. We see this as a fruitful collaboration of AI and humans, where we have gained a powerful (though not automatically reliable) friend who can think further ahead than us, and who is a better scholar than us drawing on insights from our own narrower community but also beyond. But as we already said in our introduction, we merely offer this as one group's experience, and we feel that the results and proofs presented here still need further human scrutiny, striving toward seeing the `why' of the results presented here behind the `that'. For more concrete details of the uses of AI leading to this Note, see Appendix C.

\section{Appendix A: Reducing Tiling to $\LC$}

We give an explicit reduction of Wang tiling to satisfiability of the equality-free logic of functional dependence extended by Baltag--van Benthem's conditional dependence atoms. The construction uses five fixed variables, unary predicates, singleton agreement modalities, and conditional atoms with atomic conditions. No conditional nesting is needed.
\begin{theorem}\label{CWSmain}
Satisfiability, and consequently validity, of equality-free LFD extended by the atoms in \eqref{eq:cond} is undecidable. This holds already with five fixed variables, only unary ordinary predicate symbols, and atomic conditions $\alpha$ in every conditional dependence atom.
\end{theorem}

The external undecidability input is the Wang domino problem in its currently standardly used version referring to the north-east quadrant of $\mathbb{N} \times \mathbb{N}$.
%Plane and first-quadrant tileability are equivalent for this unconstrained problem \cite[p.~178]{robinson}. 
For a finite nonempty tile set $Q$, let $R_H,R_V\subseteq Q\times Q$ specify the allowed east and north pairs. It is undecidable whether there is $\lambda:\N^2\to Q$ respecting these pairs. 

Given a tile set $Q$ where $R_H,R_V\subseteq Q\times Q$ specify its east and north matching pairs, we need to show that there is a formula $\Phi_Q$ in $\LC$ such that $Q$ tiles $\N^2$ if and only if $\Phi_Q$ is satisfied in a dependence model.

\subsection{The formula constructed from a tile system}
We first construct the formula $\Phi_Q$, whose satisfiability in a dependence model is supposed to ensure the existence of a grid-like structure in the dependence model. More explanation on why the formula is constructed as such will come in the next two subsections. Readers should not spend too much time staring at the formulas appearing in this subsection before they start reading the explanation. 

Use the five variables
\[
u,\quad e_{00},\quad e_{01},\quad e_{10},\quad e_{11}.
\]
Introduce unary predicates $C_{ij}$ for $i,j\in\{0,1\}$ and $T_\tau$ for $\tau\in Q$ in the language. 
Below $C_{ij}$ abbreviates the actual LFD atom $C_{ij}(u)$, and $T_\tau$ abbreviates $T_\tau(u)$. 
%These are ordinary unary predicate applications, not new propositional primitives.

For a finite family $\psi_1,\ldots,\psi_k$, define
\[
\one(\psi_1,\ldots,\psi_k):=
\left(\bigvee_{r=1}^k\psi_r\right)\land
\bigwedge_{r<s}\neg(\psi_r\land\psi_s).
\]
Empty disjunctions mean $\bot$. All indices $i,j,a,b$ in the following schemas range over $\{0,1\}$.

Let $\Gamma_Q$ be the conjunction of the following formulas:
\begin{align}
&\one(C_{00},C_{01},C_{10},C_{11}),\quad
\one(T_\tau:\tau\in Q); \tag{P}\label{part}\\[2pt]
&C_{ij}\to D_{\{e_{ib}\}}^{C_{1-i,j}} e_{i,1-b};
\tag{HP}\label{hp}\\[2pt]
&C_{ij}\to D_{\{e_{aj}\}}^{C_{i,1-j}} e_{1-a,j};
\tag{VP}\label{vp}\\[2pt]
&C_{ij}\to
\left(\Dia{e_{i0}}C_{1-i,j}\land\Dia{e_{0j}}C_{i,1-j}\right);
\tag{S}\label{ser}\\[2pt]
&C_{ij}\land T_\tau\to
\Boxx{e_{i0}}\left(C_{1-i,j}\to
\bigvee_{(\tau,\sigma)\in R_H}T_\sigma\right);
\tag{TH}\label{th}\\[2pt]
&C_{ij}\land T_\tau\to
\Boxx{e_{0j}}\left(C_{i,1-j}\to
\bigvee_{(\tau,\sigma)\in R_V}T_\sigma\right).
\tag{TV}\label{tv}
\end{align}
Set
\begin{equation}\label{phi}
\Phi_Q:=C_{00}\land\All\Gamma_Q.
\end{equation}
This is an effective construction. The only new dependence atoms occur in \eqref{hp} and \eqref{vp}; every superscript is a single unary atom. Both values of $b$ in \eqref{hp} and both values of $a$ in \eqref{vp} must be included.

\subsection{From a satisfying model to a tiling}
Suppose $\mathcal M,s_{00}\models\Phi_Q$. On its admissible assignments, define $sE_{ab}t$ by $s(e_{ab})=t(e_{ab})$. Each $E_{ab}$ is a genuine equivalence relation. Each assignment has a unique colour $(i,j)$ by \eqref{part}.
Define auxiliary directed relations $H,V$ on assignments by
\begin{align*}
H(s,t)&\ \Longleftrightarrow\
\bigvee_{i,j}\bigl(C_{ij}(s)\land C_{1-i,j}(t)\land sE_{i0}t\bigr),\\
V(s,t)&\ \Longleftrightarrow\
\bigvee_{i,j}\bigl(C_{ij}(s)\land C_{i,1-j}(t)\land sE_{0j}t\bigr).
\end{align*}
The relations $H,V$ are metatheoretic definitions; they are not additional symbols in $\Phi_Q$. 

We are going to extract a grid-like structure from the dependence model by following these two relations, horizontally and vertically. 
\begin{lemma}[Square propagation]\label{square}
If $H(s,t)$, $V(s,s')$, and $V(t,t')$, then $H(s',t')$.
\end{lemma}
\begin{proof}
First observe that the following relation between two assignments $s$ and $t$, which satisfy $C_{ij}$ and $C_{1-i,j}$ respectively, is ensured by \eqref{hp}:
\[
sE_{i0}t\quad\Longleftrightarrow\quad sE_{i1}t.
\]
Similarly, \eqref{vp} makes sure that 
\[
sE_{0j}t\quad\Longleftrightarrow\quad sE_{1j}t.
\]
if $s\models C_{ij}$ and $t\models C_{i,1-j}$.

Hence if $s\models C_{ij}$, $t\models C_{i-i,j}$, $s'\models C_{i,1-j}$ and $t'\models C_{1-i,1-j}$, by the above observation and the definition of the relations $H$ and $V$, we have
\[
sE_{ij}t,\qquad sE_{ij}s',\qquad tE_{ij}t'.
\]

So, $s'E_{ij}t'$ by symmetry and transitivity. When $j=0$, this is already the defining guard $s'E_{i0}t'$. When $j=1$, apply \eqref{hp} at $s'$ with $b=1$ to obtain $s'E_{i0}t'$. The target assignment $t'$ satisfies $C_{1-i,1-j}$, so $H(s',t')$.
\end{proof}

By \eqref{ser}, every assignment has an $H$ successor and a $V$ successor. Choose an infinite $H$ path $s_{00},s_{10},s_{20},\ldots$. Having chosen row $n$, choose a $V$ successor $s_{m,n+1}$ of every $s_{mn}$. Lemma~\ref{square} implies that row $n+1$ is also an $H$ path. Iteration yields
\[
H(s_{mn},s_{m+1,n}),\qquad
V(s_{mn},s_{m,n+1})\qquad(m,n\in\N).
\]
This is a homomorphism of the first-quadrant grid into the assignment structure. Distinct grid points need not map to distinct assignments.

Define $\lambda(m,n)$ to be the unique tile whose predicate is true at $s_{mn}$ which is ensured by \eqref{part}. Formula \eqref{th} yields
$(\lambda(m,n),\lambda(m+1,n))\in R_H$, and \eqref{tv} yields
$(\lambda(m,n),\lambda(m,n+1))\in R_V$. Therefore $\lambda$ is a valid tiling.

\subsection{From a tiling to a genuine dependence model}
Suppose $\lambda:\N^2\to Q$ is a valid tiling. Take the object domain to be the disjoint union
\[
O=\bigl(\{00,01,10,11\}\times\mathbb Z^2\bigr)
\ \dot\cup\ \bigl(\{u\}\times\N^2\bigr).
\]
For each grid point $(m,n)$ include the assignment $s_{mn}$ defined by
\begin{align}
s_{mn}(u)&=(u,m,n),\notag\\
s_{mn}(e_{ab})&=
\left(ab,\left\lfloor\frac{m-a}{2}\right\rfloor,
\left\lfloor\frac{n-b}{2}\right\rfloor\right).
\label{blocks}
\end{align}
Let $S=\{s_{mn}:m,n\in\N\}$. If the language has other variables, give them fixed values throughout $S$. 
Interpret $C_{ij}$ as the set of objects $(u,m,n)$ for which
$(m\bmod2,n\bmod2)=(i,j)$, and $T_\tau$ as the set of $(u,m,n)$ for which $\lambda(m,n)=\tau$. That is, 
\begin{align*}
  s_{mn}\models C_{ij}(u)\qquad &\text{iff}\qquad (m\bmod2,n\bmod2)=(i,j)\\
  s_{mn}\models T_\tau(u)\qquad &\text{iff}\qquad \lambda(m,n)=\tau\\
\end{align*}

\vspace{-4ex}

We are going to show that $\Phi_Q$ is true at $s_{00}$. By the construction, it is easy to see that \eqref{part} holds everywhere and $C_{00}$ holds at $s_{00}$. 
To show that \eqref{ser} hold everywhere, we need this lemma: 
\begin{lemma}
For $x,x'\in \N$ and $a\in \{0,1\}$, assume that 
\[
\left\lfloor\frac{x-a}{2}\right\rfloor
=\left\lfloor\frac{x'-a}{2}\right\rfloor,
\]
\begin{itemize}
  \item if $x - a \bmod 2 = 0$, then $x' = x+1$ or $x' = x$;
  \item if $x - a \bmod 2 = 1$, then $x = x' + 1$ or $x = x'$.
\end{itemize}
\end{lemma}
Taking an $s_{mn}$ which satisfies $s_{mn}\models C_{ij}$, by the above lemma, if $m - i \bmod 2 = 0$, then $s_{mn}(e_{i0}) = s_{m+1,n}(e_{i0})$ and $s_{m+1,n}\models C_{1-i,j}$; if $m - a \bmod 2 = 1$, $s_{mn}(e_{i0}) = s_{m-1,n}(e_{i0})$ and $s_{m-1,n}\models C_{1-i,j}$; the remaining case where $m = 0 $ and $i = 1$ is impossible when $s_{mn}\models C_{ij}$. So, $s_{mn}\models \Dia{e_{i0}}C_{1-i,j}$. Similarly, we can prove that $s_{mn}\models \Dia{e_{0j}}C_{i,1-j}$.

To show that \eqref{hp} and \eqref{th} holds everywhere, the key step is the same. That is, taking an arbitrary assignment $s_{mn}$ which satisfies $s_{mn}\models C_{ij}$ and an arbitrary assignment $s_{m'n'}$ which satisfies $s_{m'n'} \models C_{1-i,j}$, we must prove that if $s_{mn}(e_{ib}) = s_{m'n'}(e_{ib})$, then $m' = m+1$ and $n' = n$. 

The assumption $s_{mn}(e_{ib}) = s_{m'n'}(e_{ib})$ tells us that 
\[
\left\lfloor\frac{m-i}{2}\right\rfloor
=\left\lfloor\frac{m'-i}{2}\right\rfloor,
\qquad
\left\lfloor\frac{n-b}{2}\right\rfloor
=\left\lfloor\frac{n'-b}{2}\right\rfloor.
\]
By the above lemma, the second equation tells us that $|n - n'| = 1$ or $|n-n'|= 0$. Since $n \bmod 2 = n'\bmod 2$, we have $n = n'$. For $m$ and $m '$, 
because $s_{mn}\models C_{ij}$, by the above lemma, $m' = m$ or $m' = m+1$. Since $m' \bmod 2 = 1-i$, $m\neq m'$. So, $m'=m+1$.
We have proved the key step. 

From the conclusion that $m' = m+1$ and $n' = n$, it easily follows that $s_{mn}(e_{i,1-b}) = s_{m'n'}(e_{i,1-b})$ and thus \eqref{hp} holds at $s_{mn}$. Moreover, by the valid tiling $\lambda$, if $s_{mn} \models T_\tau$ and $(\tau,\sigma)\in R_H$,  the conclusion implies that   $\lambda(m',n') = \sigma$, that is, $s_{m'n'} \models \bigvee_{(\tau,\sigma)\in R_H} T_\sigma$. So, \eqref{th} also holds at $s_{mn}$.
\eqref{vp} and \eqref{tv} holds everywhere can be proved in a similar way. 

\section{Appendix B: Refuting Direct Definability}

We put our main results in perspective by showing that the three extended dependence logics considered in this article allow no  direct definitions in one another working on the same models.

\smallskip
Fact 1. $\LE$ is not definable in the other two systems.

\smallskip

Reason: $\LC$ and $\LI$ both have the earlier Separation Property: 
if we make the value ranges for all variables disjoint, the same formulas remain true 
at the same (simply adapted) assignments. But the formula $x = y$ is not invariant under this transformation.

\smallskip

Fact 2. $\LI$ is undefinable in $\LE$. 

\smallskip

Reason: If $\varphi $ in $\LE$ defined the atom $I_xy$, then, by the Separation
Property for $\LI$ we can assume that all subformulas $x = y$ of $\varphi$ are false. 
Therefore, $\varphi$ reduces to an equivalent $\mathsf{LFD}$-formula. But $I_Xy$ is not definable in $\mathsf{LFD}$, 
since $\mathsf{LFD}$ is decidable and $\LI$ is not.

\smallskip

Fact 3. $\LI$ is undefinable in $\LC$.

\smallskip

Reason: Consider $I_xy$, and take a dependence model for two variables
$x, y$ in that order and no predicate symbols (this suffices for undefinability
of $I_xy$ without loss of generality) with assignment set \{11, 12, 13, 21, 22, 32, 33\}. 

$I_xy$ holds at $11, 12, 13$ and nowhere else. Now consider what $\LC$ can
define here. $D_xy$ is false everywhere, and so is $D_yx$. ($D_xx$ and $D_yy$ 
are equivalent to true.) In this model, $\LC$-formulas can
only define the whole set or the empty sets of assignments. This can be shown 
by induction with key steps for $D^{\varphi}_xy, D_x\varphi$ (the other cases are symmetric).

\smallskip

Fact 4. $\LC$ is not definable in $\LE$.

\smallskip

This is true because of the Separation Property, by the same reasoning as for $\LI$.

\smallskip

Fact 5. $\LC$ is not definable in $\LI$.

\smallskip
%Consider the atom D^{Pxy}_xy. It suffices to look at LFD + I formulas having
%just x, y, P. Take a two-variable model with {11, 12, 21, 22} and P = {(1, 1)}.
%
%All D_xy, D_yx are False, D_xx, D_yy True. Pxy is true only at 11.
%
%D^{Pxy} xy is true at 11, 21, 22, but not at 12.
%
%Independence atoms I_xy, I_yx are True everywhere, I_xx, I_yy are False. 
%I_{x, y} x, I_{x, y}y False. LFD + I (plus P) can define {11}, {12, 21, 22},
%but nothing else I think. E.g., D_x Pxy is False everywhere.

Reason: Consider the setting of Fact 6.15 in \cite{BV21}, with only two variables $x, y$ and one atom $Pxy$. Consider a dependence model $\mathfrak{M}$ with two assignments $s, t$, where $s(x) = s(y) = 0, t(x) = 0, t(y) = 1$ and $P$ holds only for the pair $(0, 0)$. Now we extend $\mathfrak{M}$ to a model $\mathfrak{M'}$ with an assignment $u$ such that $u(x) = 0, u(y) = 2$, and $P$ holds of the pair $(0, 2)$. 

Consider a map $F: \mathfrak{M'}\to M, \, s, u\mapsto s, t\mapsto t$. It is not hard to see that for all $\LI$-formulas $\varphi$:  $\mathfrak{M'}, v\models\varphi$ iff $\mathfrak{M}, F(v)\models\varphi$. (In particular, all independence atoms $I_xy$, $I_yx$ are true everywhere in both models.) However, $D^{Pxy}_xy$ is true at $s$ in $\mathfrak{M}$  but false at $s$ in $\mathfrak{M'}$.

\medskip

\begin{remark} The above arguments are ad-hoc. It might be more elegant to show all these facts 
uniformly using appropriate notions of bisimulation for the three languages. \end{remark}

\section{Appendix C: A Protocol of Our Use of AI}
In this project, we used $\mathsf{ChatGPT}$ to investigate dependence logics, discuss candidate constructions, and assist with checking and presenting proofs. Both the research questions and our interaction with the AI evolved as the project progressed. We briefly describe this process in three stages, together with a few methodological observations.

The first stage began in late September 2026 with the decidability problem for \(\LC\). Our discussions with the AI yielded two proofs of undecidability: a tiling argument and a reduction from \(\LE\) to \(\LC\). After checking them, we focused mainly on reorganizing the proofs and improving their exposition. Comparing the two approaches also drew our attention to the connections between logical translations revealed by the reduction.

The second stage began in mid-October. We conjectured that the independence extension \(\LI\) could also be related to the other logics by reductions, and asked the AI to investigate its relationship with first-order logic. The resulting construction used a quotient domain, linking it to the reconstruction of a common domain in the first stage.

The third stage took place mainly in late October. We asked the AI for direct reductions from \(\LI\) to the other two extensions of \(\LFD\). However, the proposed constructions were more complex than the earlier ones. We therefore turned to a route through first-order logic and tried to generalize the method of Grädel and Pützstück. This generalization also encountered difficulties. Further questions about these difficulties led the AI to provide a direct satisfiability reduction from equality-free first-order logic to \(\LE\), which we checked and reorganized before incorporating it into the manuscript in early November.

By early November, connections between logics had become the central focus of the paper. In the present version, the faithful translations in Sections 2 and 4 yield satisfiability reductions as corollaries, alongside the separate SAT reductions in Section 3 and Appendix A.

Throughout this process, the AI assisted with exploring candidate constructions, checking individual steps, and drafting proofs. The authors checked the relevant constructions and proofs and were responsible for selecting the results, explaining the ideas, and preparing the final text. Some AI-generated results checked and confirmed by the authors were not included in the paper. The authors take responsibility for the content of this paper.

This experience shows that AI can contribute new constructions and proof ideas. The gradual shift from decidability to translations also reflects the authors’ role in comparing and interpreting the results and deciding which questions to pursue. In the third stage, the choice of methods and analysis of difficulties helped guide further inquiry. Organizing and explaining the proofs helped make the resulting mathematical connections explicit. We hope this work provides an encouraging example of human-AI collaboration in logic.
\end{document}